%% file: main.tex
\documentclass[journal]{IEEEtran}
\usepackage{subfigure}
\usepackage{color}

\ifCLASSINFOpdf
  \usepackage[pdftex]{graphicx}
  
\else
  
\fi

\usepackage{graphicx}
\usepackage{caption}

\usepackage{eurosym}
\usepackage{amsmath,amssymb,amsfonts}
\usepackage{graphicx}
\usepackage{textcomp}
\usepackage[para]{footmisc}
\usepackage{xcolor}
\usepackage{hanging}
\usepackage{graphicx}
\usepackage{float}
\usepackage{subfiles}
\usepackage{hyperref}
\usepackage[ruled,lined,vlined,linesnumbered]{algorithm2e}

\SetCommentSty{mycommfont}
\usepackage{textcomp}
\usepackage{listings}
\usepackage{multirow}
\usepackage{svg}
\usepackage{url}
\usepackage{mathptmx}
\usepackage{csquotes}

\usepackage{hyperref}
\usepackage{url}

\begin{document}

\title{Agent-based Framework for Self-Organization of Collective and Autonomous Shuttle Fleets}

\author{
Antonio~Bucchiarone\IEEEauthorrefmark{1}\href{https://orcid.org/0000-0003-1154-1382}{\includegraphics[scale=.06]{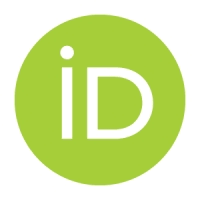}},
Martina De Sanctis\IEEEauthorrefmark{2}\href{https://orcid.org/0000-0002-9417-660X}{\includegraphics[scale=.06]{figures/orcid.png}}, 
Nelly Bencomo\IEEEauthorrefmark{3}\href{https://orcid.org/0000-0001-6895-1636}{\includegraphics[scale=.06]{figures/orcid.png}}

\vspace{0.5em}
\IEEEauthorrefmark{1}Fondazione Bruno Kessler, Trento, Italy\\
bucchiarone@fbk.eu\\[0.3em]
\IEEEauthorrefmark{2}Gran Sasso Science Institute (GSSI), L'Aquila, Italy\\
martina.desanctis@gssi.it\\[0.3em]
\IEEEauthorrefmark{3}SEA, CS, Aston  University, Birmigham, UK\\
nelly@acm.org\\[0.3em]
}

\markboth{IEEE Transactions on Intelligent Transportation Systems,~Vol.~xx, No.~yy, May~2020}%
{Shell \MakeLowercase{\textit{et al.}}: Bare Demo of IEEEtran.cls for IEEE Journals}

\maketitle

\begin{abstract}
The mobility of people is at the center of transportation planning and decision-making of the cities of the future. In order to accelerate the transition to zero-emissions and to maximize air quality benefits, smart  cities  are  prioritizing  walking, cycling, shared mobility services and public  transport over the use of private cars.  
Extensive progress has been made in autonomous and electric cars. Autonomous Vehicles (AV)  are  increasingly capable  of  moving  without  full  control  of humans, automating some aspects of driving, such as steering or braking.
For these reasons,  cities are  investing  in  the  infrastructure  and technology needed to support connected, multi-modal transit networks  that  include shared  electric Autonomous Vehicles (AV). 
The relationship between traditional public transport and new mobility services is in the spotlight and need to be rethought.   
This paper proposes an agent-based simulation framework that allows for the creation and simulation of mobility scenarios to investigate the impact of new mobility modes on a city daily life. It lets traffic planners explore the cooperative integration of AV using a decentralized control approach. A prototype has been implemented and validated with data of the city of Trento.

\end{abstract}


\IEEEpeerreviewmaketitle

\section{Introduction}
\label{sec:intro}
\input{introduction}

\section{State of the art}
\label{sec:rw}
\input{sota}

\section{System Architecture and Models}
\label{sec:models}
\input{models}

\section{Execution Layer}
\label{sec:execution}
\input{execution}

\section{Analysis Layer}
\label{sec:analysis}

The \textsc{Analysis Layer} executes experiments on top of the \textsc{Execution Layer} to evaluate properties of self-organized systems. We have specified and executed a set of experiments to evaluate the performance of the proposed solution. Specifically, we are interested to figure out the efficiency of the decentralized approach in terms of served users, waiting time, travel costs, AS usage, in different settings involving diverse number of users and/or AS. We aim to show how simulations can support the mobility service provider to infer the best size of the fleet for serving a given number of users, while guaranteeing the best costs and services for the passengers, as well as avoiding underused shuttles.

The simulations are built on top of real data of the city of Trento (Italy). 
The experiments were run on a laptop equipped with a quad-core CPU running at 2.7GHz, and 16Gb memory. In the following, we discuss three main evaluations that have been performed.

\subsection{Evaluating the approach with multiple vs. a few passengers destinations}
This evaluation aimed to understand the impact of the approach on the efficiency of passengers allocation when considering the following complementary scenarios: (i)~each user is assigned a different working place, meaning that diversified and mainly non overlapping travel destinations are considered; 
(ii)~each user is assigned one of the only two available working places; (iii)~each user is assigned the same unique working place.
The last two scenarios differ from the first one, since they model those cases in which a company can decide to adhere to the AS mobility service and offer it to its employees. This impacts on the final destination of users, used for planning their trip to work, which will coincide. Considering that users' destinations are exploited by AS to decide if get users on board or not, we aim to investigate the impact of only one or a few shared destinations.

\begin{figure}[!ht]
    \centering
\includegraphics[width=.5\textwidth]{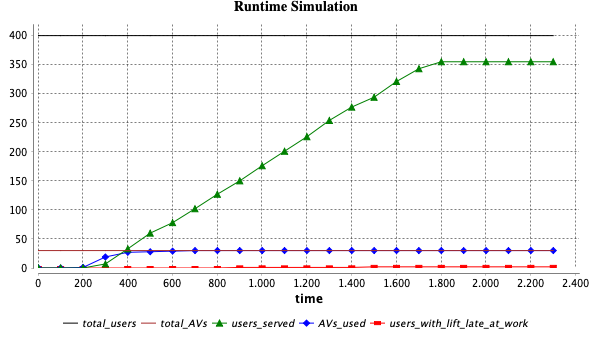}
    \caption{\textit{Exp1} with diversified working places.}
    \label{fig:exp1}
\end{figure}

\begin{figure}[!ht]
    \centering
\includegraphics[width=.5\textwidth]{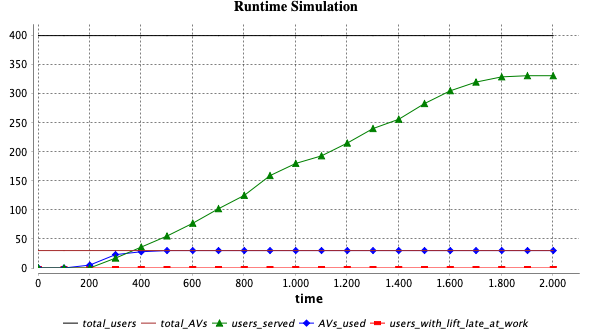}
    \caption{\textit{Exp2} with two working places.}
    \label{fig:exp2}
\end{figure}

\begin{figure}[!ht]
    \centering
\includegraphics[width=.5\textwidth]{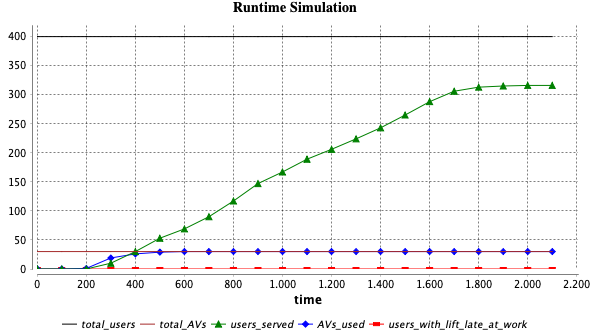}
    \caption{\textit{Exp3} with only one working place.}
    \label{fig:exp3}
\end{figure}

The initial setting for this first evaluation considers a fixed set of agents modeling a fleet of 30 shuttles of 12-places each and 400 users. The cost assumed for the service is of EURO 1 per kilometer~(Km), which is then shared among the users sharing the same travel. The time interval for the users to start working ranges between 8:00~am and 10:00~am, meaning that they do not look for a lift to work all at the same time. 
We have run three experiments for this evaluation, $Exp1$, $Exp2$ and $Exp3$, simulating the three scenarios described above, in which diversified, 
two and only one working places are considered.

In Figures~\ref{fig:exp1},~\ref{fig:exp2}, and ~\ref{fig:exp3}, we report the charts of the three experiments showing the simulation progress till its end.  Specifically, in the \emph{x-axis} the number of simulation cycles are reported, where the minimum duration of a cycle corresponds to 0.01 seconds. Then, each chart displays the number of users served by the fleet of AS, on top of the 400 users requesting for a lift, the total number of AS used and the number of served users arriving late at work. At first sight, it can be noticed that the number of served users slightly decreases while moving from $Exp1$ to $Exp3$. 

In Table~\ref{tab:comparisonTable}, further details about the first evaluation are reported. In particular, first column of Table~\ref{tab:comparisonTable} lists the most interesting outcomes that have been measured during the simulation, and which are grouped as follows: in the upper part of the table (from the second to the fifth row) there are parameters referring to the users; in the lower part of the table (last three rows) there are parameters referring to the AS. Several observations can be made by looking at the table. As already seen from Figures~\ref{fig:exp1},~\ref{fig:exp2}, and ~\ref{fig:exp3}, the number of served users slightly decreases from 355 in $Exp1$, to 331 in $Exp2$ and 316 in $Exp3$, while only in $Exp1$ we registered two served users arrived late at work. 

Interestingly, in the first scenario, where each user is assigned a different working place, thus diversified destinations are considered ($Exp1$), we can observe a higher average travel cost for users, namely EURO 2,60 w.r.t. EURO 1,50 and EURO 1,45 in $Exp2$ and $Exp3$ respectively, and a considerably lower average waiting time of 2,61 minutes, against the 5,44 and 7,18 minutes in $Exp2$ and $Exp3$, respectively. The higher average travel cost is supported by the Kms traveled by the AS, on average, that in $Exp1$ is roughly the double w.r.t. $Exp2$ and $Exp3$. 
Lastly, we also measured the cumulative user lifts per each AS, i.e., the total number of users served by an AS at the end of its work. There is not a significant difference on the cumulative user lifts given on average by each AS, in the three scenarios, which is only slightly greater in $Exp1$ where, however, more users are served. Considering that it ranges roughly between 7 and 17 in the scenarios, we can also state that a fleet of 30 shuttles is enough for a total of 400 users. 

\begin{table}[htb!]
\centering
\caption{Comparison between \textit{random} ($Exp1$) and \textit{fixed} ($Exp2$ and $Exp3$) number of working places.}
\label{tab:comparisonTable}
\resizebox{\columnwidth}{!}{
\begin{tabular}{lccc}
\cline{2-4}
\multicolumn{1}{l|}{} & \multicolumn{1}{c|}{\textbf{$Exp1$}} & \multicolumn{1}{c|}{\textbf{$Exp2$}} & \multicolumn{1}{c|}{\textbf{$Exp3$}} 
\\ \cline{2-4}  \hline 
\multicolumn{1}{|l|}{\textbf{Users served}} & \multicolumn{1}{c|}{355} & \multicolumn{1}{c|}{331} & \multicolumn{1}{c|}{316} 
\\ \hline
\multicolumn{1}{|l|}{\textbf{Served users late at work}} & \multicolumn{1}{c|}{2} & \multicolumn{1}{c|}{0} & \multicolumn{1}{c|}{0} 
\\ \hline
\multicolumn{1}{|l|}{\textbf{Avg. travel costs}} & \multicolumn{1}{c|}{EURO 2,60 } & \multicolumn{1}{c|}{EURO 1,50 } & \multicolumn{1}{c|}{EURO 1,45 }  
\\ \hline
 \multicolumn{1}{|l|}{\textbf{Avg. waiting time}} & \multicolumn{1}{c|}{2,61 min} & \multicolumn{1}{c|}{5,44 min} & \multicolumn{1}{c|}{7,18 min}  
\\ \hline \hline
\multicolumn{1}{|l|}{\textbf{Avg. cumulative user lifts [range]}} & \multicolumn{1}{c|}{12 {[}8-17{]}} & \multicolumn{1}{c|}{11 {[}7-16{]}} & \multicolumn{1}{c|}{11 {[}7-15{]}} 
\\ \hline
\multicolumn{1}{|l|}{\textbf{Tot. gain for the AS service}} & \multicolumn{1}{c|}{EURO 923,95 } & \multicolumn{1}{c|}{EURO 497,04 } & \multicolumn{1}{c|}{EURO 457,62 } 
\\ \hline
\multicolumn{1}{|l|}{\textbf{Avg. Km traveled by AS}} & \multicolumn{1}{c|}{29,42 km} & \multicolumn{1}{c|}{16,60 km} & \multicolumn{1}{c|}{15,25 km}  
\\ \hline 
 & \multicolumn{1}{l}{} & \multicolumn{1}{l}{} & \multicolumn{1}{l}{} 
 \\
 & \multicolumn{1}{l}{} & \multicolumn{1}{l}{} & \multicolumn{1}{l}{} 
\end{tabular}
}
\end{table}

\begin{figure}[!h]
\vspace{-0.5cm}
    \centering
\includegraphics[width=255pt]{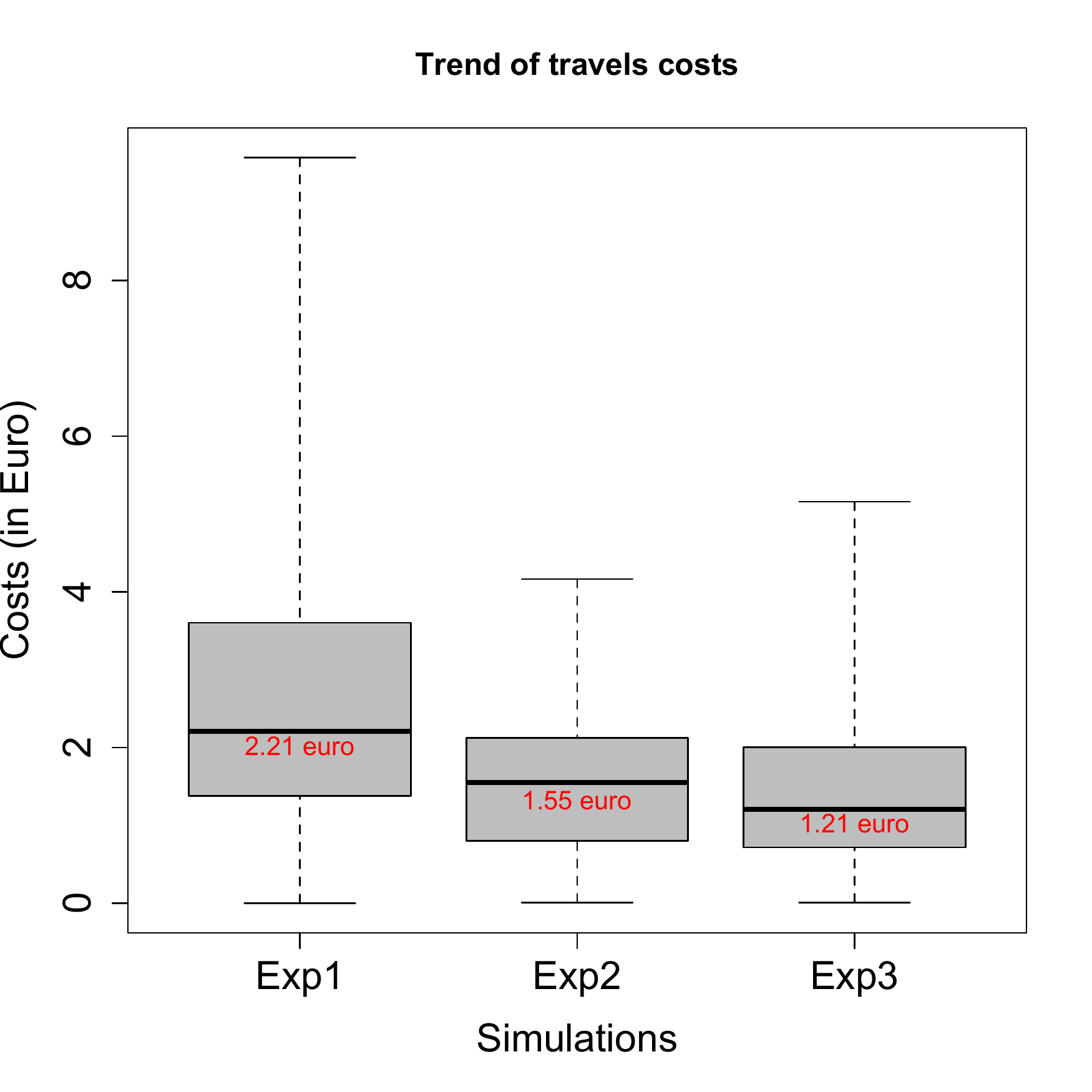}
\vspace{-0.2cm}
    \caption{Trend of travel costs in the three experiments. }
    \label{fig:costsFirstSimulation}
\end{figure}

Figure~\ref{fig:costsFirstSimulation} shows the distribution of travel costs in the three scenarios, where the median value is reported. It highlights a wider distribution of costs in the case in which diversified working places are considered, bringing to the higher average travel cost discussed in Table~\ref{tab:comparisonTable}. From this evaluation, it results that managing a considerable set of users sharing the same working place contributes to making the AS service less expensive and thus, more convenient for companies who may at the same time wish to offer it to their employees, at the expense of increase of waiting time. 

However, $Exp1$ shows that when diversified working places are considered, travel costs are subject to a high variance that might discourage the use of a shared ride. This requires the need of keeping the travel costs bounded, in such a way to make the approach reliable. Considering that the travel cost mainly depends on the distance AS cover (see equations~(\ref{eq:costLeg}) and~(\ref{eq:costPath})), at a first stage we are considering to bound such distance. In particular, the high variance shown in Figure~\ref{fig:costsFirstSimulation} depends on the way in which AS identify their first group of passengers among those who made a lift request, as described in Algorithm~1. AS compute their initial path based on the passengers destinations, by selecting the farthermost away as their final target. 
Our idea for future work is that of modifying our approach by computing an upper bound to the final target distance from the origin of AS, by gradually excluding the passenger whose destination is the farthermost away but overpasses the established upper bound. Eventually, since travel costs are also affected by the number of passengers sharing a ride (see equation~(\ref{eq:costLegPass})), we plan to further extend our approach in order to maximize shared route, as done in~\cite{8062833}.

\subsection{Evaluating the efficiency of a fleet of a fixed number of AS}
\begin{figure*}[htb!]
\begin{minipage}[t]{\textwidth}
\begin{minipage}[]{0.5\textwidth}
       \begin{center}
        \includegraphics[width=\textwidth, keepaspectratio]{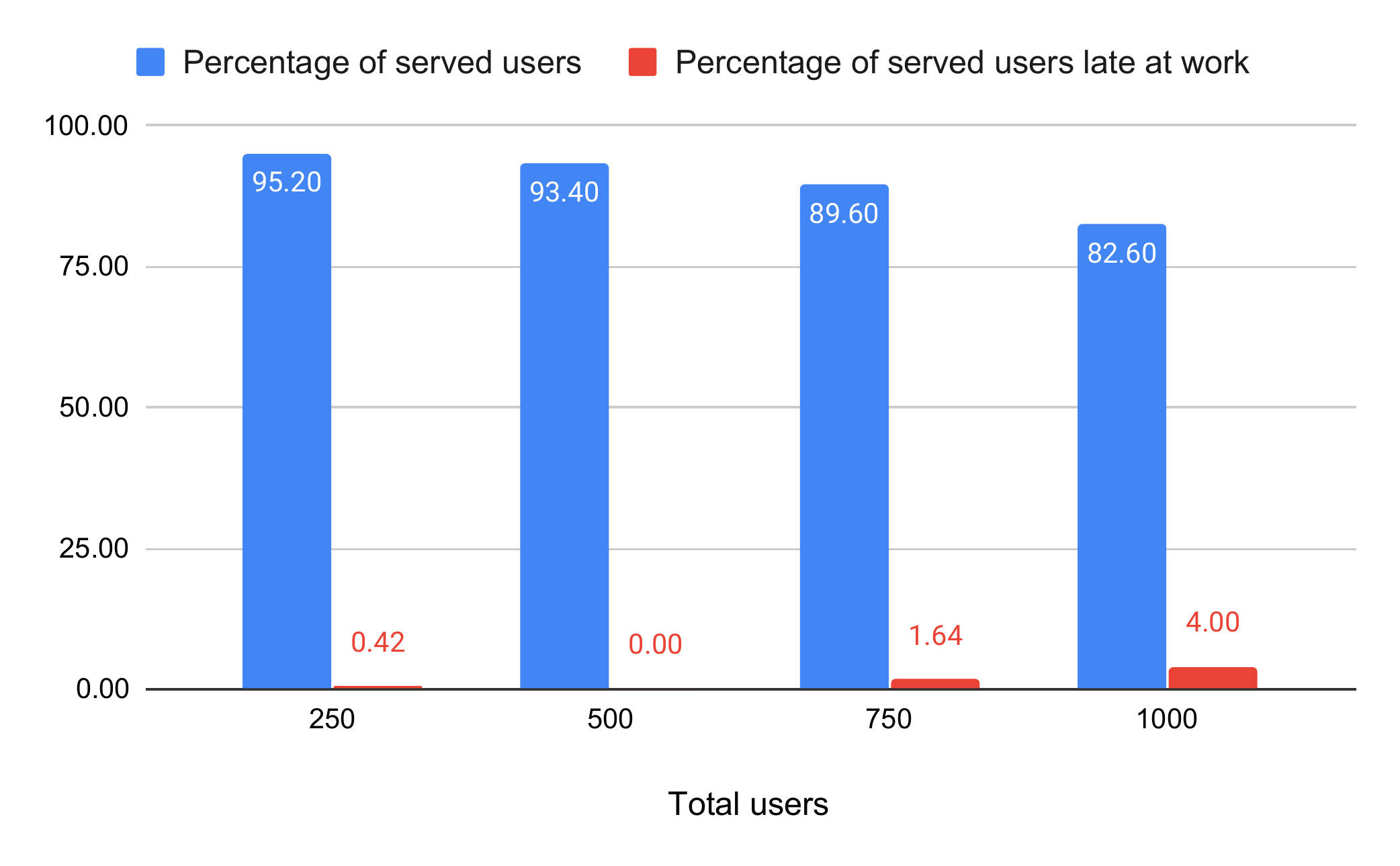} 
        \end{center}
\end{minipage}
\hspace{0.2cm}
\begin{minipage}[]{0.5\textwidth}
    \begin{center}
     \includegraphics[width=\textwidth, keepaspectratio]{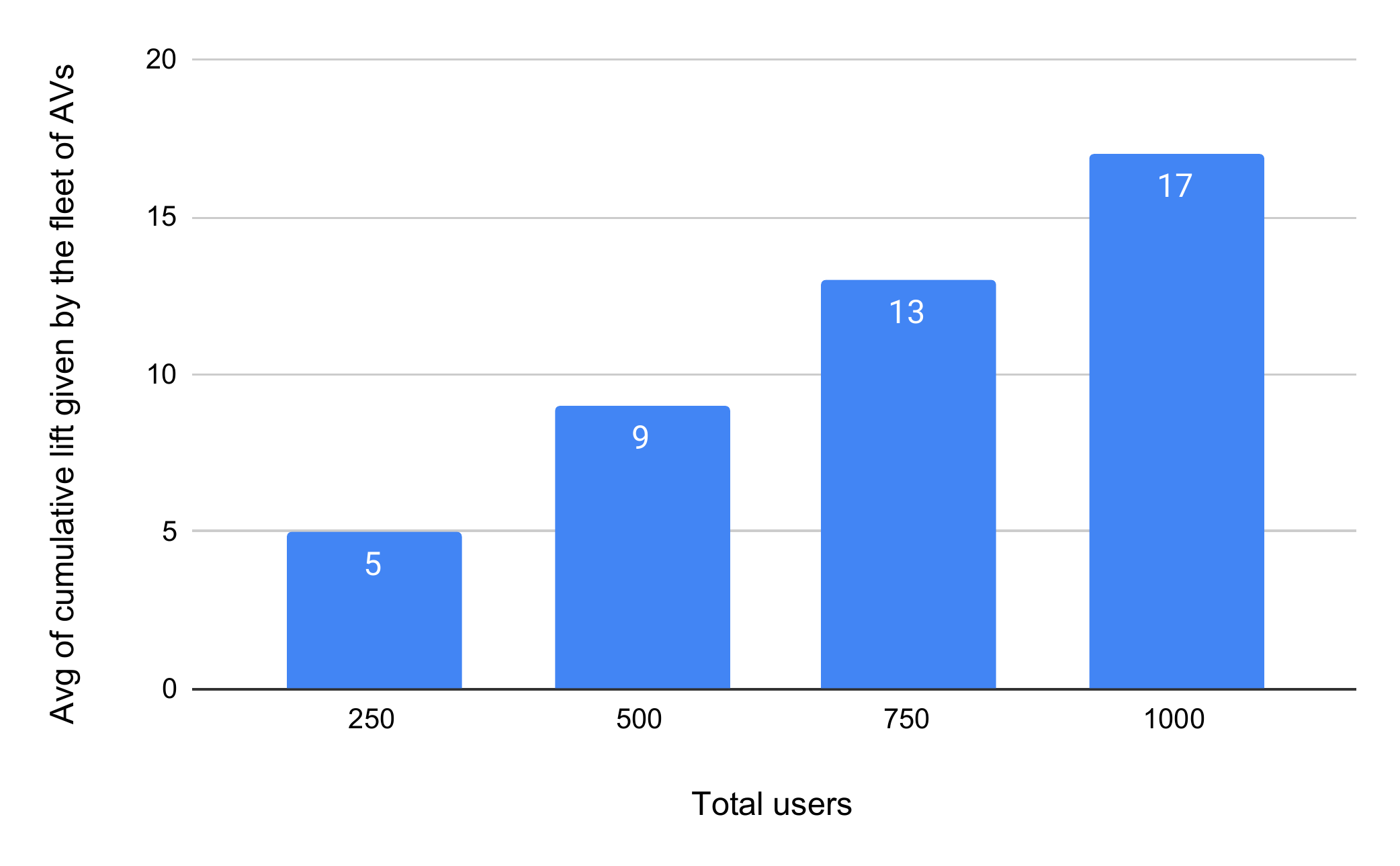}
    \end{center}
\end{minipage}
\end{minipage}
\caption{Fixed fleet of 50 shuttles in scenarios with varying number of total users, from 250 to 1000. \textbf{(Left side)} Percentage of served users over the total number of users and percentage of users late at work among the served users; \textbf{(Right side)} Average of cumulative user lifts given by the fleet in the different scenarios with varying number of users.} \label{fig:secondSimulation} 
\end{figure*}

This evaluation aimed to measure the efficiency of a fixed fleet made by 50 shuttles of 12-places each, with varying numbers of total users, namely 250, 500, 750 and 1000. As in the first evaluation, the cost for the service is of EURO 1 per Km whereas the time interval for the users to start working ranges between 8:00 am and 10:00 am. In Figure~\ref{fig:secondSimulation} left side we show the percentage of served users (per each users total number) as well as the percentage of served users arriving late at work. At the same time, for each simulation we further measured the average of cumulative user lifts given by the AS in the fleet, in Figure~\ref{fig:secondSimulation} right side, in order to understand to what extend AS reach or surpass their full occupancy capacity. 

It can be observed that, in terms of served users, the fleet of 50 AS guarantees good performances in all the four scenarios considered, from the 95,20\% of served users over a total of 250 users, to the 82,60\% over 1000 users. In contrast, the percentage of served users arriving late at work has incremented as the total number of users requesting lifts increases, even if the number of users arriving late at work remains significantly low.
The most relevant numbers are registered in the case of 750 and 1000 users, where 1,64\% (i.e., 11 users) and 4,00\% (i.e., 33 users) among served users arrive late at work, with an average late time of 6,11 and 6,70 minutes, respectively.
However, to have a complete overview of the performance of the AS fleet, it is important to observe the average number of cumulative user lifts given by the AS, in each scenario. Indeed, from Figure~\ref{fig:secondSimulation} right side it results that, when serving 250 and 500 users, AS do not reach their full occupancy capacity (i.e., 5 and 9 cumulative user lifts in average) and this might negatively impact on the travel cost split among users sharing the same travel. 
To the contrary, when serving 750 and 1000 users, the registered average of cumulative user lifts is of 13 and 17, respectively, representing an optimized use of the entire fleet. Eventually, we measured the waiting time for users whose request for a lift has been accepted, in the four analyzed scenarios. The average waiting time increases with the amount of the number of total users. Specifically, it ranges from 1,23 minutes with 250 users to 3,31 minutes in the scenario with 1000 users.

In conclusion, this second evaluation shows that a fleet of 50 AS achieves a good trade-off between the percentage of served users and the usage of AS when serving a number of users in between 750 and 1000, while it is more efficient in terms of served users but, it is also less efficient in terms of AS usage, with less than 750 users approximately.

\subsection{Evaluating the efficiency of the approach in finding the best size of a AS fleet for serving a given amount of users}
\begin{figure*}[t!]
\begin{minipage}[t]{\textwidth}
\begin{minipage}[]{0.5\textwidth}
       \begin{center}
       \centering
        \includegraphics[width=\textwidth, keepaspectratio]{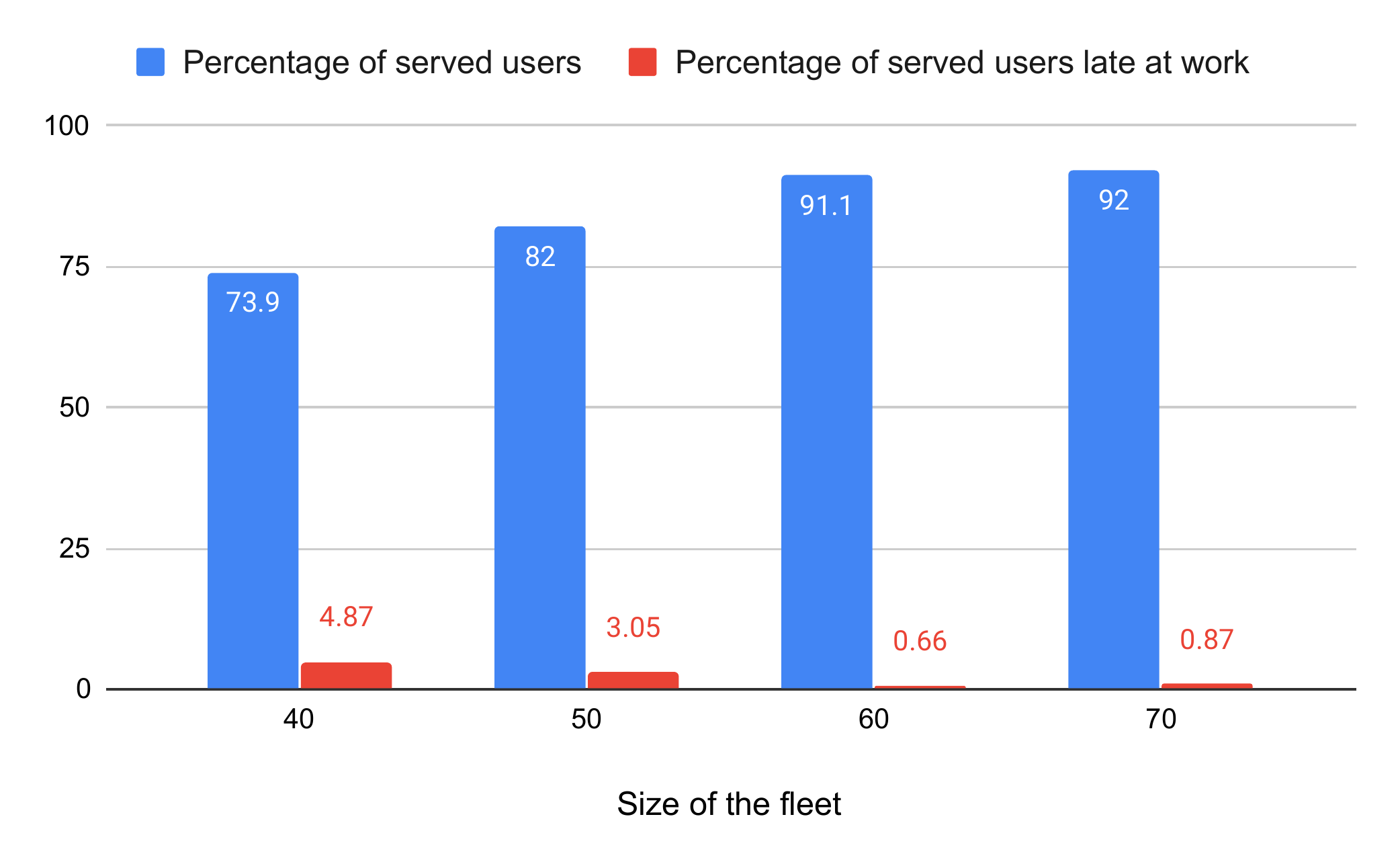}
        \end{center}
\end{minipage}
\hspace{0.2cm}
\begin{minipage}[]{0.5\textwidth}
    \begin{center}
    \centering
     \includegraphics[width=\textwidth, keepaspectratio]{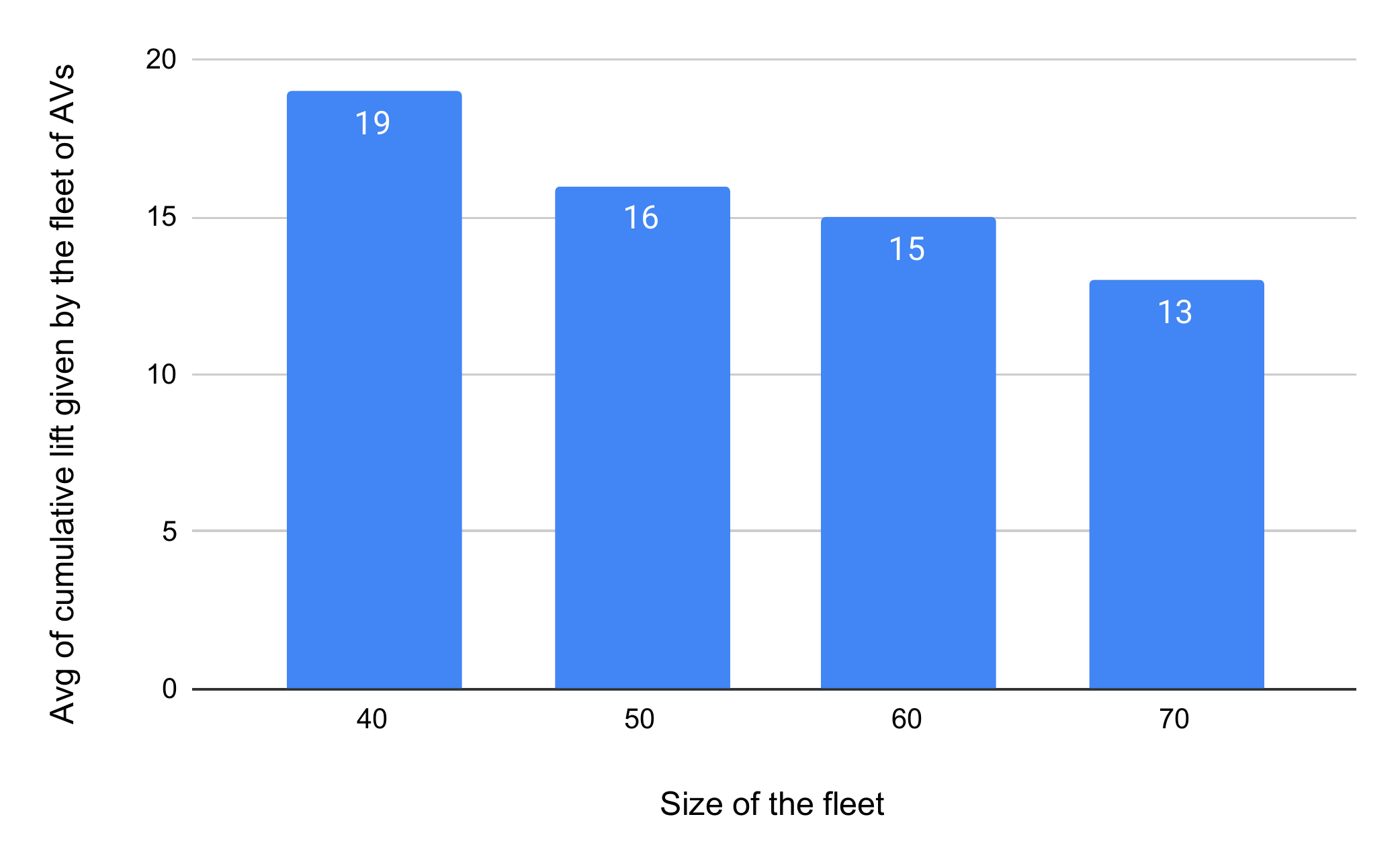}
    \end{center}
\end{minipage}
\end{minipage}
\caption{Fixed number of 1000 users in scenarios with varying number of AS in the fleet, from 40 to 70. \textbf{(Left side)} Percentage of served users over the total users and percentage of users late at work among served users; \textbf{(Right side)} Average of cumulative user lifts given by the fleet in the different scenarios with varying number of AS.}
\label{fig:thirdSimulation} 
\end{figure*}

The third evaluation aimed to establish the best size of a AS fleet for serving a fixed amount of 1000 users, by running simulations with a fleet of different size, i.e., made by 40, 50, 60 and 70 AS. The basic setting is the same as in the first and second evaluations. 
In Figure~\ref{fig:thirdSimulation} left side we show the percentage of served users per each fleet size as well as the percentage of served users arriving late at work. At the same time, for each fleet we measured the average of cumulative user lifts given by the AS, in Figure~\ref{fig:thirdSimulation} right side, to observe the trend of AS occupancy capacity.

As expected, the left side of Figure~\ref{fig:thirdSimulation} shows that the number of served users increases with the increasing size of the fleet and it ranges from 73,9\% (when the fleet is made by 40 AS) to 92\% (with a fleet of 70 AS). On the other hand, the number of users served who arrived late at work decreases with the increase of the size of the fleet. It varies from 4,87\% (i.e.,~36 users) over the served users, for which we registered an average late time of 6,36 minutes, to 0,87\% (i.e., 8 users) of late users with an average late time of 2,7 minutes.
Instead, the right side of Figure~\ref{fig:thirdSimulation} shows how fleets of different size perform in terms of AS occupancy capacity, given the fixed number of 1000 users. We can observe that, in all scenarios, on average the number of cumulative user lifts provided by the AS surpass the number of their occupancy capacity, i.e., 12, ranging from 13 to 19.  Of course, the average of cumulative user lifts is higher when the fleet is made by 40 AS, and it slowly decreases when the size of the fleet rises. This basically shows a good usage of the entire fleet. Further, it allows us to predict that increasing the fleet size over 70 might not be convenient, since it can lead to an under-utilisation of the fleet, increasing the number of AS that do not reach their full occupancy capacity. Moreover, as we have seen, a fleet of 70 shuttles served the 92\% of 1000 users, which can be considered as a good result.  This simulation further shows that a fleet of 60 AS presents as well a good performance in terms of both served users (91.1\%) and AS utilization (15 cumulative user lifts on average). The latter will be seen as convenient or not based on the objective of the service provider. 

Lastly, the waiting time for users served by the AS, decreases with the increasing size of the fleet, from 4,01 minutes whit a fleet of 40 AS, to 2,13 minutes whit a fleet of 70 AS.

\begin{figure*}[htb!]
\begin{minipage}[t]{\textwidth}
\begin{minipage}[]{0.5\textwidth}
       \begin{center}
       \centering
        \includegraphics[width=0.99\textwidth, keepaspectratio]{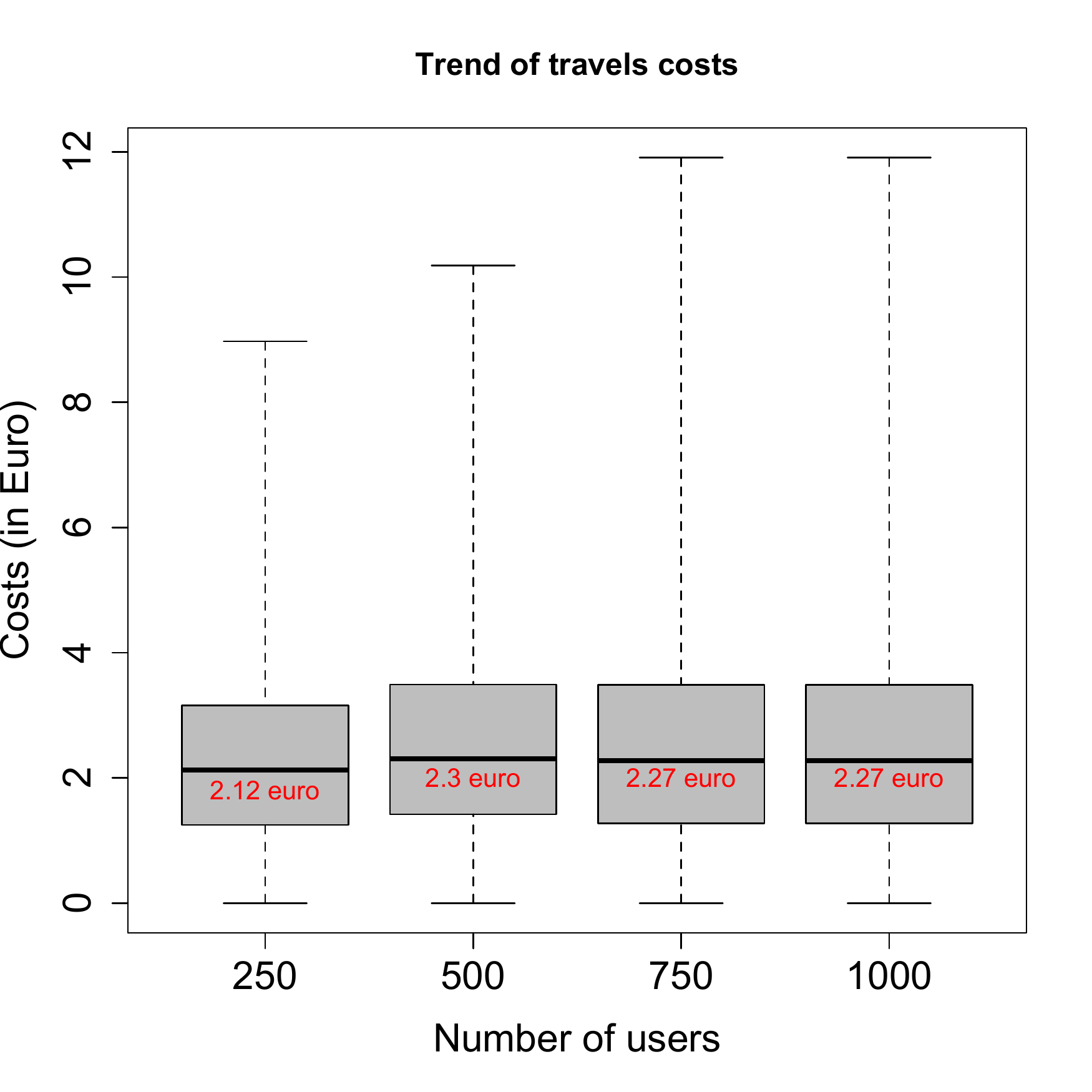}
        \end{center}
\end{minipage}
\hspace{0.2cm}
\begin{minipage}[]{0.5\textwidth}
    \begin{center}
    \centering
     \includegraphics[width=0.99\textwidth, keepaspectratio]{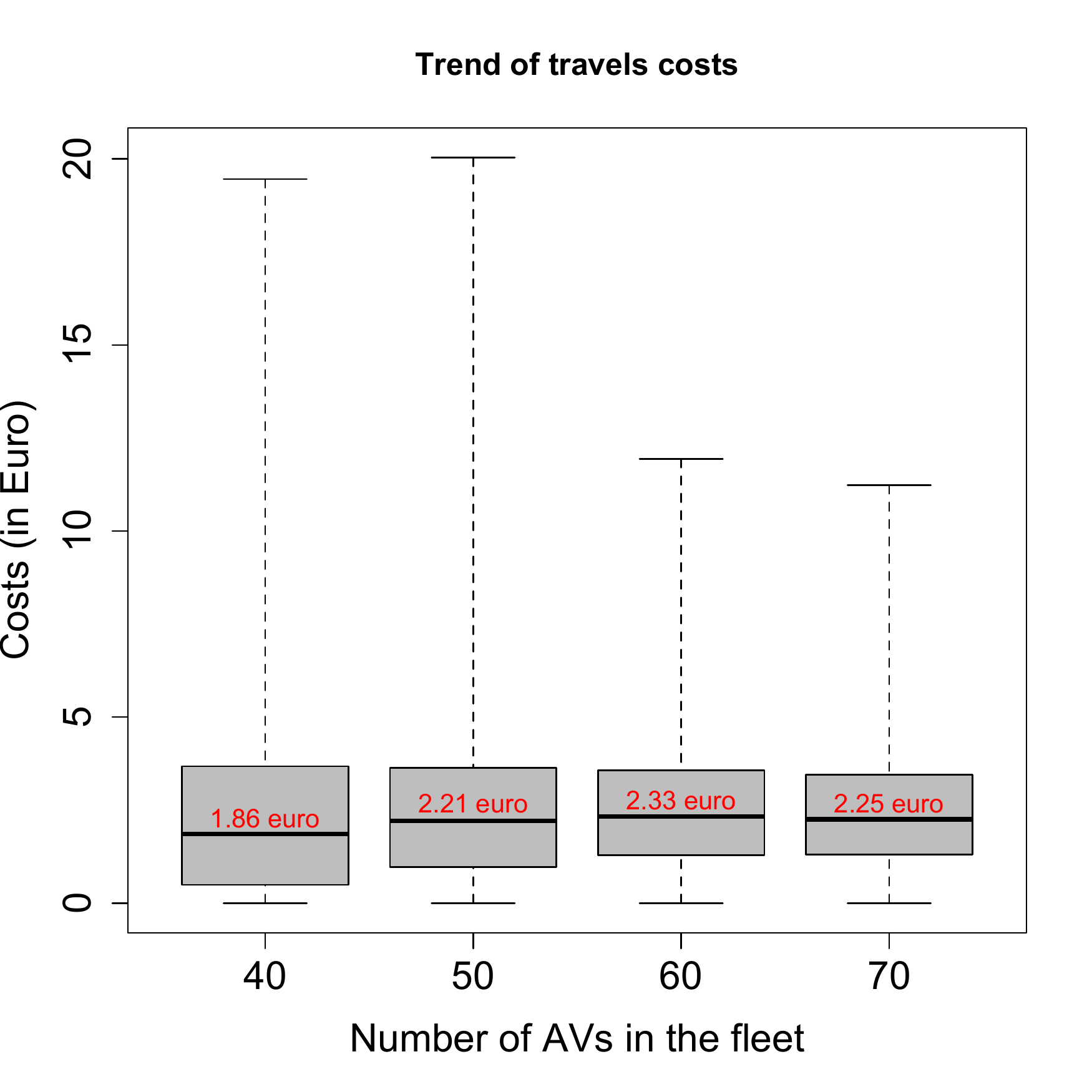}
    \end{center}
\end{minipage}
\end{minipage}
\caption{Trend of travel costs. \textbf{(Left side)} Scenario with fixed fleet of 50 AS and varying number of total users; \textbf{(Right side)} Fixed number of 1000 users and varying number of AS in the fleet.}
\label{fig:coststSecondAndThirdSim} 
\vspace{-0.3cm}
\end{figure*}

At last, to have a complete overview of the cost for the service provided by the AS, in Figure~\ref{fig:coststSecondAndThirdSim} we plotted the travel costs for the served users in both the second evaluation with the fixed fleet of 50 AS and varying number of total users (left side) and the third evaluation with the fixed number of 1000 users and varying size of the fleet (right side). The median value is reported in red. 
We can see that, on average, the cost of the service is stable in all the scenarios studied, which it is of about EURO 2 for the majority of served users. However, there are cases in which the travel is more expensive, reaching peaks of roughly EURO 12 and EURO 20. This can be due to longer distances covered by the AS but also to a low level of occupancy capacity of AS, which leads to the situation where the travel cost is shared among just few users.

\paragraph{Threats to validity} Our experimentation may be internally biased from the settings of input parameters, namely the \textit{time interval} for the users to start working, the \textit{cost} per Km, the \textit{angle} between the passengers destinations and the AS's current location used to decide if accepting or not passengers on the road. Variations to these parameters could affect travel time and costs, and the number of passengers being late at work, whereas the overall procedure is not affected. We have chosen input parameters based on real working settings we experience in our daily life, although we are aware that multiple combinations of them are possible. 
As external threats to validity, we are aware that the application of the approach to other case studies, i.e., different cities, has not been performed. This might lead to observe different results due to, e.g., the diverse dimensions of cities, with different road graphs structure, but we leave this point for our future work.

\subsection{Discussion}

From the described evaluations above, we can draw several conclusions. The first evaluation highlights two different perspectives in which the mobility service providing AS fleets can be exploited. The first refers to the perspective of users, where each user decides if adhering or not to service, based on their working place or routines. The second instead, gives the perspective of companies that may decide to adhere to the AS mobility service in order to offer it to their employees. For instance, a company can decide to support collective mobility for environmental regarding matters. Of course, these two alternatives are not complementary but they can be used simultaneously in order to perform trade-offs. We have performed a preliminary evaluation of these alternatives, and analysed in isolation, and discovered that the scenario in which a consistent amount of users work for the same company (i.e., in the same working place) leads to lower travel costs and to a lower amount of Kms traveled by the AS, at the expense of the users waiting time, which even if tending to increase they also remain within acceptable levels. 
The second and third evaluations instead, show how the proposed agent-based framework can support the service provider in delivering useful insights on the number of users to be served by a fleet, and the size of a fleet serving a given amount of users that should be adopted. Further simulations allowed us to evaluate up to which extent a fleet can guarantee good performance in terms of served users, while also keeping low the number of users reaching their working place late, and the cumulative user lifts per AS as higher as possible. 

All the previous evaluations allow us to conclude with the following thoughts. A fundamental aspect of a smart and sustainable city is to be able to sense the pulse of the city, to perform short and long-term analysis of phenomena and to provide valuable information to decision makers. 
Simulations are relevant in this context as they support administrators, operators and users in the assessment of how innovations and proposed solutions for mobility systems will meet their needs to help plan for the future. 
With the framework proposed in this paper~\footnote{The source of the framework is available for its usage at \url{https://bit.ly/30SdGTa}. It contains: (1) the \textsc{GAMA Project} and the \textsc{MODELS descriptions}; these can be further used by the research community to model other transportation modes and simulate their impact on cities. (2) the \textsc{SCRIPTS} to create the Trento city map and for decentralised simulations, and (3) the installation, documentation and usage instructions.}, we provide a solution that is able to: (1) exploit the data collected by the different simulations, (2) enable a deep analysis of the impact of AS in a city in different scenarios and, (3) help administrators, companies and citizens understand their city and its traffic, and how it reacts when novel mobility solutions are taken into account.

\section{Conclusions and Future Work}
\label{sec:conclusions}

\input{conclusion}

\section*{Acknowledgment}
We would like to thank Annalisa Congiu, who started the work presented in this paper in her Master's degree thesis. The work has been partially sponsored by The Lerverhulme Trust Grant  RF-2019-548/9 and EPSRC Grant EP/T017627/1.
\bibliographystyle{ieeetr}
\bibliography{biblio.bib}


 \begin{IEEEbiography}
 [{\includegraphics[width=1in,height=1.25in,clip,keepaspectratio]{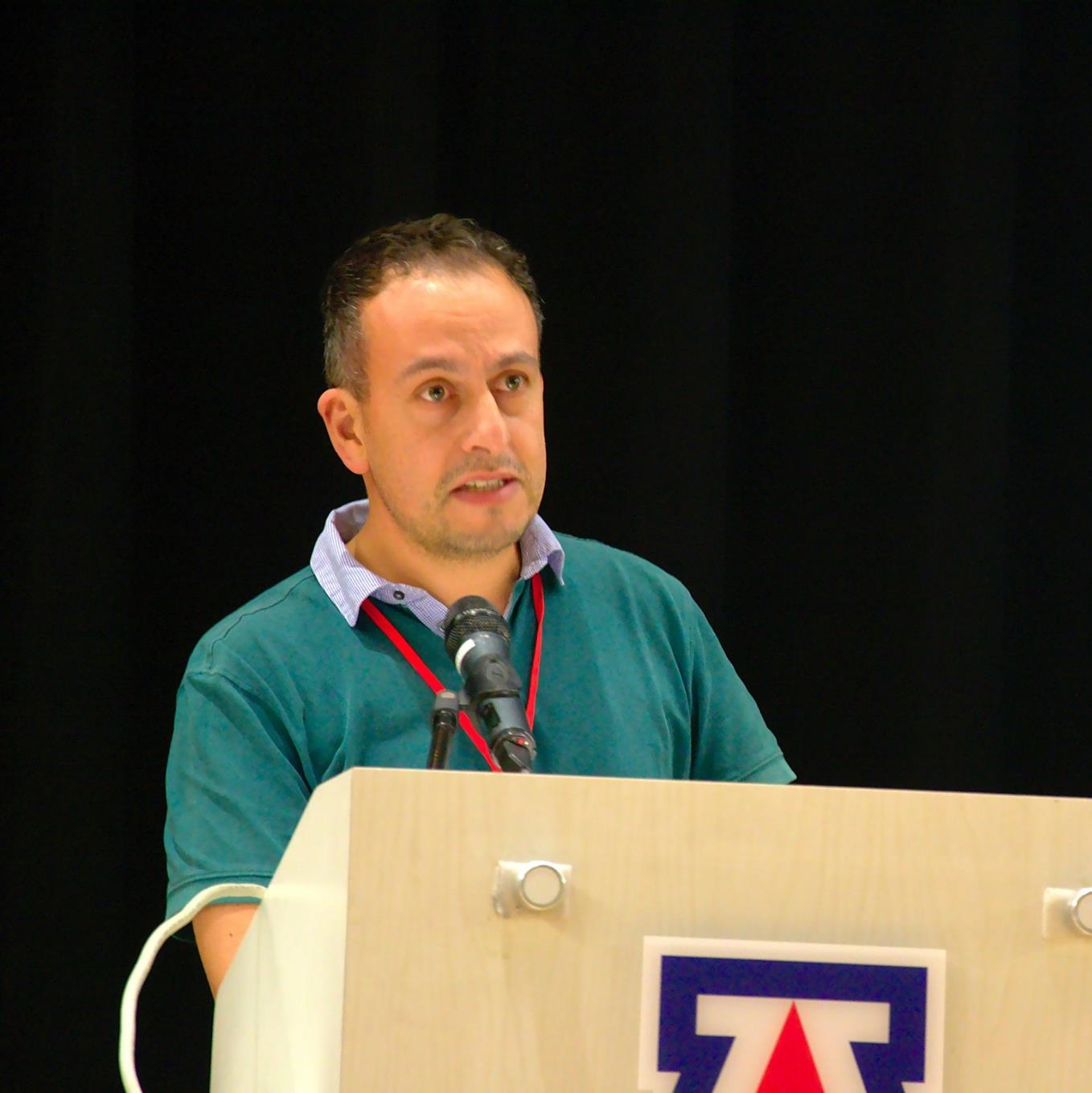}}]{Antonio Bucchiarone} is Senior Researcher within the DAS Research Unit at Fondazione Bruno Kessler (FBK) of Trento, Italy. His main research interests include: Self-Adaptive (Collective) Systems, Domain Specific Languages for Socio-Technical System, and AI planning techniques for Automatic and Runtime Service Composition. He received a Ph.D. in Computer Science and Engineering from the IMT School for Advanced Studies Lucca in 2008 and since 2004 he has been a collaborator of Formal Methods and Tools Group at ISTI-CNR of Pisa (Italy). He has been actively involved in various European research projects in the field of Self-Adaptive Systems, Smart Mobility and Constructions and Service-Oriented Computing. He was the general chair of the 12th IEEE International Conference on Self-Adaptive and Self Organizing Systems (SASO 2018) and he is an Associate Editor of the IEEE Transactions on Intelligent Transportation Systems (T-ITS) Journal, the IEEE Software Journal and the IEEE Technology and Society Magazine.
  \end{IEEEbiography}
  
  \begin{IEEEbiography}
 [{\includegraphics[width=1in,height=1.2in,clip,keepaspectratio]{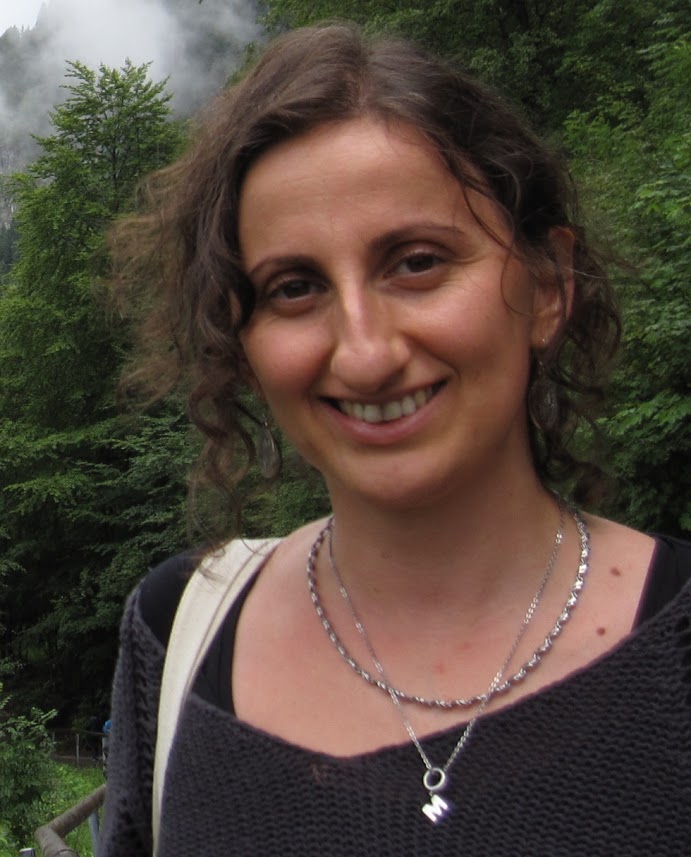}}]{Martina De Sanctis} is Assistant Professor at the Computer Science department of the Gran Sasso Science Institute (GSSI), in L'Aquila, Italy, where she has previously been Postdoctoral researcher. Her research interests include behavioral and architectural adaptation of service- and IoT-based systems, collective aspects and modeling of multi-agent systems, dynamic adaptations and its application to several domains, i.e., mobility, smart cities, IoT, eHealth. She received a Ph.D. in Computer Science at the Doctoral School in Information and Communication Technology (2018), from the University of Trento and Fondazione Bruno Kessler (FBK) in Trento. From 2013 to 2018 she was PhD fellow and researcher at FBK, at the Distributed Adaptive Systems research unit where she was working on approaches for the dynamic adaptation of service-based systems, with focus on automated service composition, and their application in different domains. During her PhD studies she actively participated in European Projects in the large-scale collective systems (ICT-FET Proactive project) and digital industry (EIT Digital project) sectors. She has been previously working in companies as software developer in the business sectors of Geographic Information Systems (GIS) and Web-based software applications.
 \end{IEEEbiography}

  \begin{IEEEbiography}
 [{\includegraphics[width=1in,height=1.2in,clip,keepaspectratio]{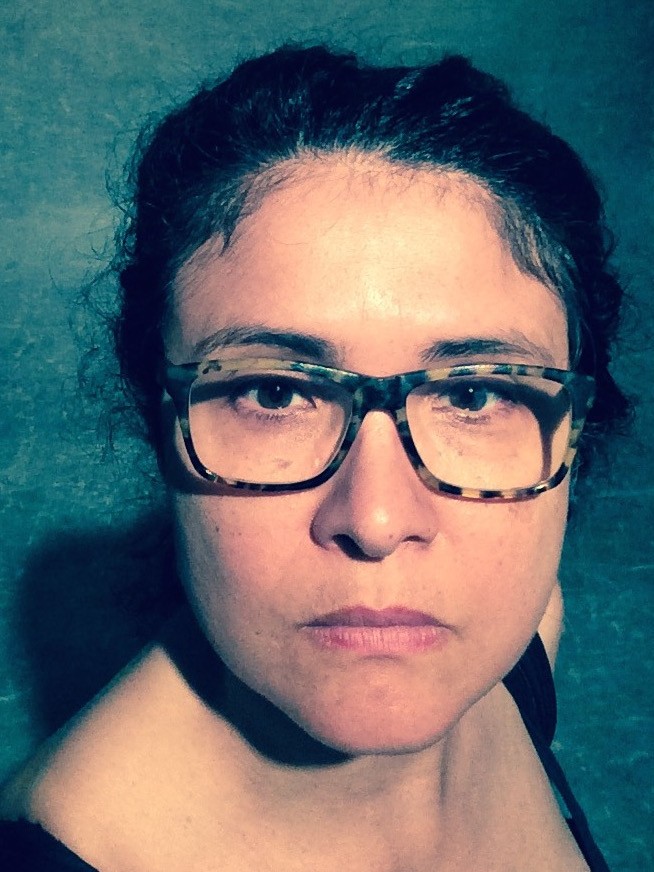}}]{Nelly Bencomo} is a Senior Lecturer in Computer Science at Aston University in the UK (since May 2013). In 2019, she was granted  the Leverhulme Fellowship "QuantUn: quantification of uncertainty using Bayesian surprises." Nelly is the principal investigator of the research project Twenty20Insight funded under the EPSRCs to work on Software Engineering, RE, and AI (2020-2023). Before, she was an EU Marie Curie Fellow, from May 2011-May 2013 under a Marie-Curie Fellowship (Grant) Requirements@run.time: Requirements-aware Systems. She was a Senior Researcher at Lancaster University until May 2011 after being was awarded her Ph.D. in Computer Science by Lancaster University in 2008. Nelly exploits the interdisciplinary aspects of software engineering, comprising both technical and human concerns while developing techniques for intelligent, autonomous and highly-distributed systems. With other colleagues, she coined the research topics models@run.time and requirements@run.time. Nelly has actively participated in different European Projects and the EPSRC in the UK in the area of self-adaptive and autonomous systems. She was the program chair of the 9th International Symposium on Software Engineering for Adaptive and Self-Managing Systems (SEAMS) in 2014, and co-program chair of the 12th IEEE International Conference on Self-Adaptive and Self-Organizing Systems (SASO) in 2018. Nelly is an Associate Editor of IEEE Transactions on Software Engineering (TSE) and a member of the  Editorial Board of the Journal of Software and Systems Modeling.

 \end{IEEEbiography}

\end{document}

%% file: introduction.tex
Transportation planning is challenging due to the uncertainty  and complexity of caring for all feasible settings and corresponding adaptation actions \cite{Lyons2016}. Further, there is urgent need for mobility innovations to be able to meet the changing needs of users. Recent progress in simulation platforms and their integration with Geographical Information Systems (GIS) make it possible to create mobility planning simulations without requiring extensive expertise.  
Urban environments with details of traffic flows are now  available from providers such as Google~\cite{Google}. Agent-based simulations allow planners to ``populate" a city or larger regions with a collection of agents that have travel patterns. Further, it is now possible to take into consideration Autonomous Vehicles (AV) in such dynamic environments \cite{Mariani2020,JeonRSLWA18}.

On the other hand, the impacts of self-driving cars portend significant changes to the transportation ecosystem\footnote{\url{http://tiny.cc/2f61pz}}. Some forecast the end of parking spaces \cite{litman2017autonomous}. Others believe that AV will paradoxically increase traffic~\cite{calvert2017will,Mehr2018CanTP}. Others predict that there will be new classes of traffic problems that occur at scale~\cite{litman2017autonomous}, due to the homogeneity of these transportation systems. 
While companies are already providing and testing the first AV~\cite{Dixit}, we are still far from knowing their impact on communities, as there is significant uncertainty concerning AV actual costs and benefits~\cite{DoRouhaniMiranda-Moreno2019,ChenJin2017}. We argue that it is necessary to develop mechanisms that allow city planners to study how AV are likely to affect travel demands and planning design space~\cite{secondSEAMSRoadmap} and decisions such as optimal road, parking and public transit supply~\cite{Litman2015AutonomousVI}, and to model, analyse, and present possible configurations in ways that the citizens can understand and participate~\cite{Gerostathopoulos2019}.

Real scenarios present different challenges. A problem we take into account covers real-time dynamic ride-sharing. It consists of sharing a vehicle among individual travellers for a trip, splitting the travel costs, i.e., tolls and parking fees, thus reducing the costs for the system in terms of cars used, pollution, traffic, etc. Such transportation allows us to reduce the costs at the expense of convenience. Nevertheless, it is not always possible to consider this form of transportation as a valid alternative due to complexity, with the addition that it is generally a disorganised and informal activity. 
Indeed, it is not simple to coordinate the itineraries of groups of travellers with different origins and destinations, or to match constrains of travellers, drivers, and AV. The complex uncertain dynamics of an environment such as a city and the need of a real-time (last-minute) approach add further levels of complexity to the riders matching.

From this perspective, the growth of \textit{autonomous shuttles} (AS) in urban public environments could enable new services to  deal  with  the  new  challenges  posed  by  large  cities, which  require  the  combination  of  the  mobility  of  people \cite{14}. In particular, several pilot experimentations prove significant technology development results, as well as the citizens’ acceptance in many cities all over the world, in countries  such  as  Germany,  France,  Switzerland,  Finland, Sweden,  The  Netherlands,  and  Estonia,  as  presented  by recent research \cite{14,15}.

This paper presents a framework for city planners to model the introduction of innovative mobility scenarios using AS, and offers the algorithms to simulate several phenomena including the decentralized and participatory management of AS. The goal of the framework is to support the decision-making of the city mobility planners in order to (i)~investigate the impact of innovative mobility modes on the traffic of a city, and (ii)~elicit information about the environment and uncertainty, with respect to unexpected situations such as breakdowns, heavy traffic and congestion when the scenarios are applied. 
The framework is multi-agent with decentralized control, and models a city (buildings, roads and intersections) with its inhabitants and cars to explore the benefits and challenges of the integration of AS into the traffic of the city. 
A main novelty aspect of our contribution is in the intersection between a decentralized collective approach and the participation of AS as agents of the collective of people, cars and elements of the city, to study the impacts of innovative mobility solutions on traffic. The paper also presents the prototype that has been implemented  and  validated  with  data  of  the  city  of  Trento.

The paper is structured as follows: Section~\ref{sec:rw} presents the state of the art. The architecture and the models for self-organized systems in the mobility domain are described in Section~\ref{sec:models}. Section~\ref{sec:execution} introduces the algorithmic solution while validation results of the prototype are discussed in Section \ref{sec:analysis}. Section \ref{sec:conclusions} concludes and discusses future work.

%% file: sota.tex
AV are vehicles that can move without the control of humans. In 2018 the U.S. Department of Transportation has released a policy regarding automated vehicles and their safe integration in the transportation system \cite{national2018us}. 
Some levels of automation have already been integrated into cars that are on the market and many manufactures are venturing in this direction \cite{bierstedt2014effects}\footnote{\url{https://bit.ly/36gZ7JC}}.

There are many advantages brought by AV, i.e., reduction of carbon production, traffic, and congestion but also fewer accidents, which are caused by driver errors, fatigue, alcohol or drugs up to a big extent~\cite{bierstedt2014effects,fagnant2014preparing,bagloee2016autonomous}. 
However, even though the levels of security provided by such system are expected to further improve and 
different test on urban roads have already been made \cite{broggi2015proud,poczter2014googles}, high level of penetration of completely autonomous vehicles cannot yet be guaranteed in the next decade \cite{national2018us,fagnant2014preparing}. As such, simulation is paramount to evaluate the AV impacts and their cooperative adaption~\cite{DoRouhaniMiranda-Moreno2019}.

In \cite{morando2018studying} the impact of AV on safety is studied using VISSIM's car following models \cite{vissim} to simulate both the human-driven vehicles and AV. Together with VISSIM, the Surrogate Safety Assessment Model (SSAM)~\cite{gettman2008surrogate} has been introduced to assess potential conflicts of a road network to be used during the simulation. The results proved that with a high penetration rates, the AV improve safety significantly.
Other studies have focused directly on modelling systems to automating intersection crossing, 
thus to help in improving the traffic flow and to minimise collisions. 
An example can be found in \cite{buzachis2018secure}, where it is shown how intersections are regulated by a Multi-Agent Autonomous Management (MA-AIM) system.
Other studies \cite{fagnant2014preparing,levin2017general,fagnant2018dynamic} focused on the application of AV to ride-sharing, to therefore study the waiting times of passengers and the impact of dynamic ride-sharing on travellers costs and the average service time in such applications. 
In \cite{sanchez2016co}, the authors focused on the decrease of carpooling users due to concerns regarding privacy and lack of trust towards other users (i.e. carpoolers). As a solution, the authors propose a decentralised manager network for carpooling, coupled with a reputation management protocol to help building trust.
In \cite{Gerostathopoulos2019} the authors present a self-adaptive framework, called TRAPP, which relies on the SUMO traffic simulator \cite{SUMO,DBLP:conf/itsc/LopezBBEFHLRWW18}. The aim is to optimize traffic flows in a decentralized and participatory way. SUMO is further exploited, together with reinforcement learning, in \cite{DBLP:conf/cosit/LiebigS17}. The goal of the approach is to avoid traffic jams with dynamic self-organizing trip planning. The impact of connected and autonomous vehicles (CAVs) on the traffic flow has been investigated in \cite{ye2018modeling}, where the vehicle-to-vehicle connection via short-range communication is also considered.

As discussed above, the scientific literature has already investigated the potential impacts of AV use on the traffic systems in cities. Expectedly, AV have potential to improve efficiency and traffic conditions in cities~\cite{QiongLu2019}. However, no study has focused on the impact of the use of \textit{Autonomous Shuttles} (AS). 
Autonomous shuttles provide an attractive and flexible solution to move people around in given areas such as industrial campuses and city centres, connecting those areas with main mass transit systems. AS can also be seen to offer new  mobility/delivery  services  into the  city center where narrow streets are not easily served by traditional buses. AS can also serve critical areas with minimal new infrastructure while reducing noise and pollution \cite{33,ASaaS2020}. The collaboration and interactions between different services that AS can provide and the needs of different entities of the city, such as neighbourhoods or employees of companies, need to be studied. Similar to some publications described above, a decentralized participatory mechanism is offered, yet we make emphasis on a decentralized agent-based collective of people, cars and elements of the city (e.g., buildings, roads).

To close this research gap identified, the focus of this paper is to realize a framework to \textit{model}, \textit{simulate} and \textit{analyze} dynamic and self-organized mobility systems through a Multi-Agent System (MAS), in which heterogeneous ensembles (AS and travellers) \cite{HolzlKPWZ15,ZambonelliBCLP11}  are created. The aim is to provide a solution to allow the simulation and analysis of different mobility scenarios in journeys, in which passengers exploit AS to reach their destinations.

%% file: models.tex
\begin{figure}[!ht]
	\centering
    \includegraphics[width=0.47\textwidth]{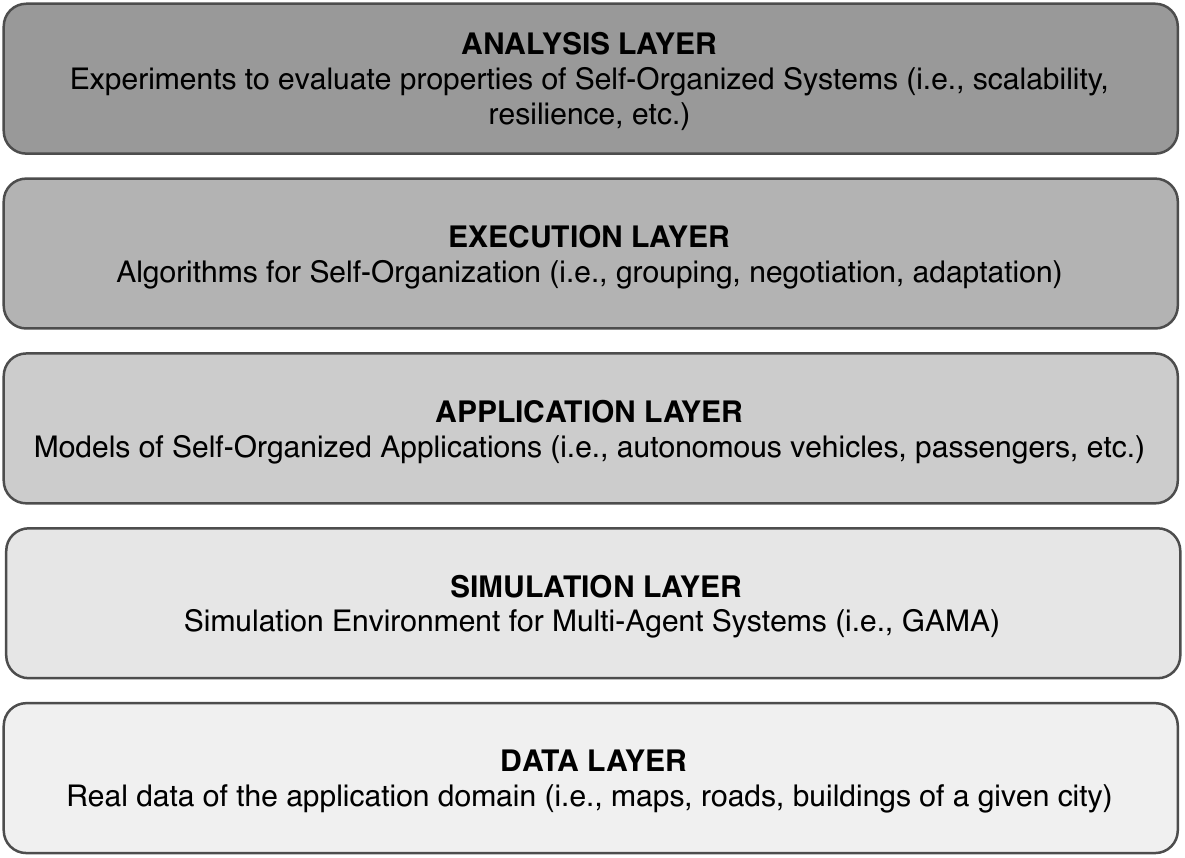}
	\caption{\label{fig:approach} Multi-Layer Architecture for Self-Organized Systems.}
\end{figure}

To provide a modular solution that can be customized for different self-organized applications in the mobility domain, we have defined a system architecture consisting of multiple layers, as shown in Figure \ref{fig:approach}, each one defined using the services of the previous ones. The system architecture and models are an extension of the work presented in~\cite{CongiuThesis}. While the initial work was focused only on autonomous vehicles and on a small portion of the Trento city map, here we consider a novel application scenario that involves AS to handle a much larger number of passengers spread over the complete map of the city. This led us to extend the architecture, the models, and to define and experiment a new distributed solution. We have also re-positioned the work w.r.t. the state of the art.

In the following sections we describe each layer with the aim of giving details on how our solution works.

\subsection{Data Layer}
The \textsc{Data Layer} is responsible for creating a virtual environment that replicates a realistic urban scenario. It has access to real data about \textit{buildings}, \textit{roads} and \textit{intersections} of a specific city. Specifically, we use OpenStreetMap\footnote{\url{https://www.openstreetmap.org}} as the framework for gathering geodata. Geodata refers to information about geographic locations that is stored in a format that can be used with any geographic information system. 

Roads and intersections provide the basis for generating a \textit{road graph}, which support the agents within the simulation. 
Next, we discuss these concepts and how they integrate into the simulated transportation network.

Buildings are polygons of diverse types stored as a map of \texttt{key:value} pairs with the type of building as value. In the GAMA framework \footnote{\url{http://www.gama-platform.org}}, buildings are grouped into two categories, namely \textit{residential} and \textit{industrial}.

Roads are polyline composed of different segments. Roads come with attributes that describe characteristics, such as the driving side, or the number of lanes.
Roads are implemented as agents as well, with basic behavior. In particular, the attribute \texttt{speed\_coefficient} helps to model traffic on the roads by slowing down the speed of shuttles in streets with high density. 
Traffic is also computed.

Crossings, traffic signals, traffic roundabouts, stops and yield signals are tagged as \texttt{highway}, along with bus stops and street lamps. In our approach, traffic signals and other relevant intersections were represented as the type \texttt{intersection} in compliance with the codes created by Patric Taillandier \cite{GAMA-traffic}. Intersections are also modeled as agents.

\subsection{Simulation Layer}
At the \textsc{Simulation Layer}, we use the  GAMA modelling and simulation framework. GAMA enables the modeling and simulation of spatially explicit agent-based systems where geodata about real-world maps, streets, buildings, etc. are integrated using GIS data. 
In GAMA, different types of agents can be programmed with their own \textit{behavior} and \textit{attributes}. The behavior of each agent is supported by functions, in the form of  \textit{reflexes} (called automatically at each time step) or \textit{actions} (called by an instance of species).
Agents can have \textit{skills}, i.e., built-in modules with a set of skill-related attributes and actions.
GAMA provides an IDE and a language, the GAma Modelling Language~(GAML). It is an agent-oriented programming language that enables users to define agents, as \textit{species}, whose behaviour is defined by actions and reflexes. 

Behind the concepts and the operational semantics of GAML, there is a meta-model made by three main categories of abstract classes, namely \textit{Entities} (i.e., agent and population of agents), \textit{Space}, \textit{Time} and \textit{Species} that combines all the previous classes.

The concept of \textit{species} is similar to that of \textit{class} in object-oriented programming. It defines the attributes, characteristics and functions of the entity it represents. 
In our approach, three species of agents have been modelled: \textit{people} who are looking for a lift to commute to work and back, \textit{autonomous  shuttles}~(AS) that offer lifts, and \textit{common cars}, such as traditional non-autonomous cars with drivers. In addition, GAMA supports the development of multi-level agent-based models~\cite{GAMAsimulator} through a \textit{containment} relationship, thus enabling the definition of \textit{macro-species} and \textit{micro-species}. 
A \textit{micro-species} can be nested inside a species, i.e., a \textit{macro-species}. Hence, it is possible for a species to host a population of micro-species. 
There may be several reasons for a multi-level representation of agents. 
For instance, there may be need to consider agents that are part of organizations and therefore to stay at different levels. 
In our approach, this relation was useful in order to represent people that enter into a shuttle by defining a micro-species inside the AS species.

Since the subject of this study is in the mobility domain, among the available built-in skills we adopted the following:
\begin{itemize}
    \item \textit{Movement skills}: Implement actions to simulate the movement of agents in an open space or along a graph. It is employed by the \textit{people} and the \textit{passenger} species.
   \item \textit{Road skills} \cite{GAMA-traffic}: Register agents on the road and is used by the \textit{road} species.
    \item \textit{Advanced Driving skill} \cite{GAMA-traffic}: Simulates agents capable of driving. Both, \textit{cars} and \textit{AS} species have this skill.
\end{itemize}

The GAMA framework also provides an interactive visualization tool, which gives feedback to the user during the simulation. 
It also allows users to automatically instantiate agents and create realistic models and simulation of a given geographic area. 

\subsection{Application Layer}
The \textsc{Application Layer} specializes the simulation layer and supports the modeling of dedicated self-organized applications. In this layer we can model the behaviors of all the agents involved in the target application.
In our specific case, \texttt{People} in the application are individuals who search for both a lift to their working place before a given time, and a lift at a given time to return home. Such agents are modelled by the species \textit{people}. Each agent is initialized with a \textit{living place} and a \textit{working place}, a \textit{starting} and an \textit{ending} working hour and a \textit{list of intersections} at a given distance from its location, used by AS to recognize people that are looking for a lift. 
This species has the \textit{moving} skill to move on different roads topologies. It can access the attributes \textit{starting point}, \textit{destination point}, \textit{current\_path} between these two points, \textit{current\_edge}, that is the current agent's location, and the \textit{speed} on the path.

To model people behaviours, two actions are defined, \texttt{search\_path} and \texttt{move}. The former simulates the search for a path in the road graph from the current location, constrained by topology. 
This is analogous to what happens in real life as nowadays, and before commuting to work, it is common to search the best option on, e.g., Google maps as there may be traffic and taking the usual route may lead to arriving late. 
Information about the time required to reach the destination (\textit{time\_to\_cover}) and the distance the agent will cover (\textit{distance\_to\_cover}) are stored, and the cost for the agent to move is computed. It keeps track of the time needed by the agent to arrive to its destination and the distance it covered. It also computes the distance the agent covered (\textit{dist\_covered\_alone}).
Two further variables, \textit{late} and \textit{actual\_time\_in} are used to track whether the agent was late at work, and the time the agent started working.

\begin{figure}[!ht]
    \centering
    \includegraphics[width=0.5\textwidth]{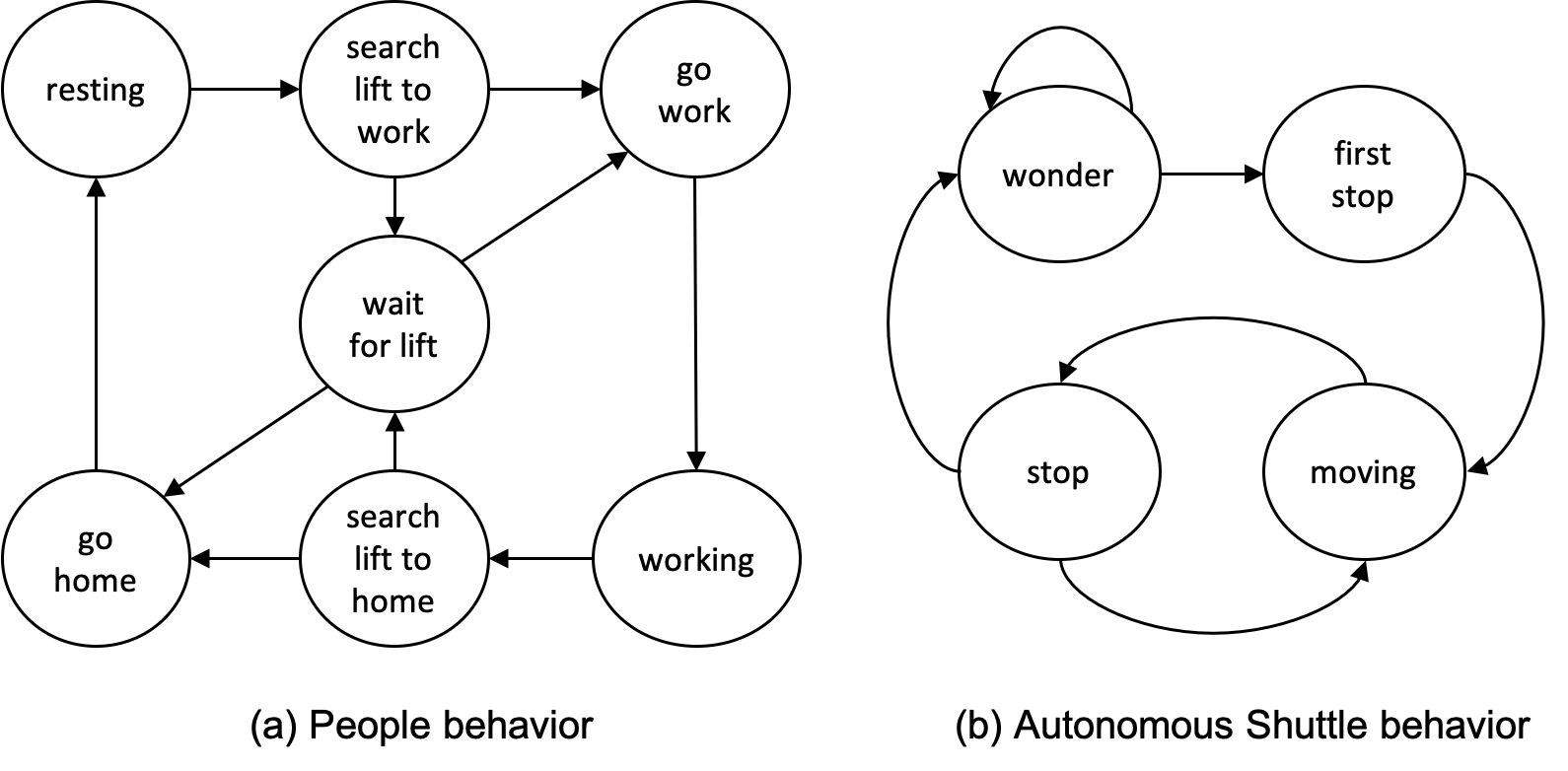}
    \caption{FSM for the People and Autonomous Shuttle behaviors.}
    \label{fig:fsm}
\end{figure}

Eventually, the agents' behavior can also be represented as a Finite State Machine~(FSM) describing the different states in which the agent can be and the transitions among them, as represented in Figure \ref{fig:fsm} left side, and further described below:
\begin{itemize}
    \item \textbf{resting}: at a given hour before the start of working time, the agent will switch its state to the state \textit{search\_lift\_to\_work}.
    \item \textbf{search\_lift\_to\_work}: at this state, the agent will set its target to the working place location, search for a route to reach it and then wait for an established amount of time for an available shuttle nearby that could pick it up. After this interval of time has passed, the agent changes its state to \textit{go\_work} if no shuttle is found, or to \textit{wait\_for\_lift} if a shuttle has been found.
    \item \textbf{wait\_for\_lift}: if the shuttle that can give a lift to the passenger is still far away, the agent will wait in this state for the shuttle to arrive.
    \item \textbf{go\_work}: this state can represent two different situations: (i)~there are not available shuttles, the agent moves alone towards its working place (the previous state is \textit{search\_lift\_to\_work}); (ii)~the shuttle it was waiting for arrives, the agent starts its lift (the previous state is \textit{wait\_for\_lift}). The agent will change its state to \textit{working} as soon as its location corresponds to its working place. 
    \item \textbf{working}: the agent is working. When it is time for the agent to go home, it will change to the state \textit{search\_lift\_to\_home}.
    \item \textbf{search\_lift\_to\_home}: after work, 
    the agent will set its target to its house location and search for the best route to return home. Then it will wait a given amount of time for a potential lift from a shuttle that is passing nearby. After this interval of time has passed, the agent changes its state to \textit{go\_home} if no shuttle is found, or to \textit{wait\_for\_lift} if a shuttle has been found.
    \item \textbf{go\_home}: after the given amount of time established to wait for a shuttle, or when the found shuttle arrives, the agent moves towards its home. Eventually, it will change its state to \textit{resting} as soon as its position corresponds to its living place.
   
\end{itemize}

The described FSM represents an abstraction of the overall people agents behaviour, which includes all the required states for our approach. Each state implicitly includes relevant parameters, such as timing concerns (e.g., in state \textit{search\_lift\_to\_work}), even if these are not visible in the FSM representation. Moreover, the FSM could be further specialized to include other parameters, if required by the analysis that is intended to be performed.

The \texttt{Driving Model} is based on the one proposed by \cite{tranouez2012multiagent} and re-elaborated by \cite{GAMA-traffic} using the GAMA platform. The agents compute a path given an origin and a destination, which may also be chosen randomly, on the road network. This trajectory is composed of a sequence of edges.
In the model presented in \cite{tranouez2012multiagent}, the movements on an edge were inspired by the Intelligent Driver Model presented in \cite{kesting2007general}, with the addition of the possibility for the drivers to change lane at any time, both when entering a new edge or when already on it. Additional properties define  maximum speed, vehicle length, or the \enquote{personality} of the driver, such as their attitude towards following the driving rules.  
In our system, two species of vehicles were simulated: a \textit{common} car and an \textit{autonomous} shuttle agents. Various instances of the former are generated in order to simulate traffic on the roads by letting them wander on the roads. For this reason, common cars have two reflexes: \texttt{time\_to\_go} and \texttt{move}. The former is used to set a random target for the agent towards which to move, the latter makes the agent move exploiting the action \texttt{drive}. The latter also checks whether the agent has slowed down to a speed below 5 kms/hour in order to decide whether or not to turn around and change route.

Figure \ref{fig:fsm} right side shows the FSM model of the AS behavior.
All the AS start at the \textit{wander} state. They will change state from \textit{wander} to \textit{first\_stop} when they find a person or a group of people searching for a lift on the road, triggering the decision-making mechanism. In this case, a final destination is chosen among the targets of the first person or group of people found in order to set an \textit{objective for the shuttle}. Once this objective is set, the shuttle can create a path proposing the related costs to the first set of probable passengers. After the first set of passengers, who accepted the shuttle offer to get on board, the agent changes its state to \textit{moving}. If there are still open seats, it may propose a lift to passengers found on the road, by changing its state to \textit{stop}, and therefore updating the costs for all passengers. 

AS keep track of the distance and the expected time needed to reach their next destination. Data also includes the time that was actually necessary to reach their destinations, the passengers on board and all the stops made to either take on or drop off passengers.

The next two layers, \textsc{Execution} and \textsc{Analysis}, are on top of the previous ones shown above, and are described next.

%% file: execution.tex
In this Section, we propose a decentralized approach for the self-organization of fleets of autonomous shuttles, as part of the \textsc{Execution Layer}.
The computational power for the grouping of agents and the computation of travel costs are distributed among the AS. Each AS is capable of dynamically finding passengers on the road it is currently travelling and automatically check whether to offer them a lift, based on certain conditions. 
\begin{figure}[!ht]
\vspace{-0.3cm}
	\centering
    \includegraphics[width=0.5\textwidth]{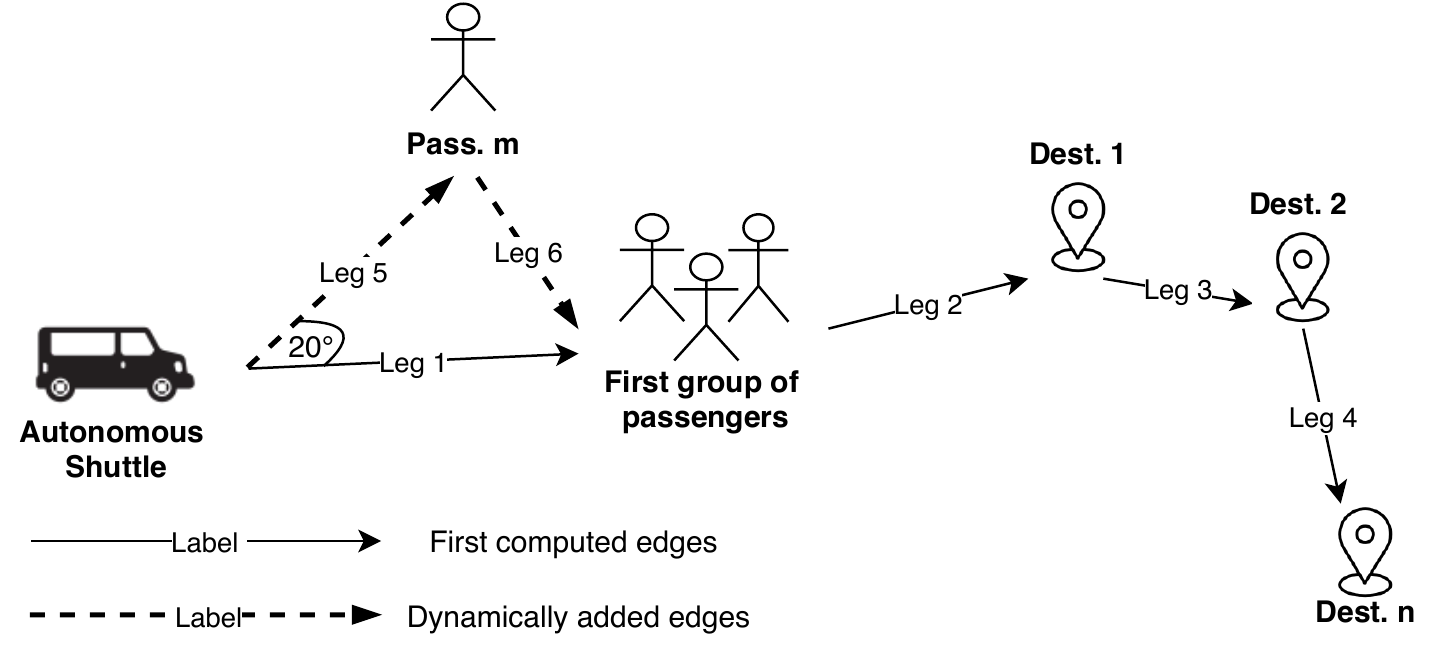}
	\caption{Dynamic addition of new passengers and path update.}
	\label{fig:pathLegs}
\end{figure}
For instance, if their initial positions and/or final destinations are too far from the current path of the AS, offering them a lift might lead to an increase of travel time and cost for those passengers already on board, which would be taken into account when deciding to accept the lift or not. 

Once started, AS identify their first passenger or group of passengers among those who made a lift request. When the first passengers are found, the AS compute an initial path based on their current positions and destinations.
A graphical representation is given in Figure~\ref{fig:pathLegs}, where the \texttt{first group of passengers} is identified. The initial path made by their position and destinations is composed by \texttt{leg~1, leg~2, leg~3} and \texttt{leg~4}.
The final target of the AS travel is computed on the base of the passenger whose destination is the farthermost away (i.e., \texttt{Dest.~n} in Figure~\ref{fig:pathLegs}). This would also increase the possibility of finding and adding more passengers on the road (i.e., \texttt{Pass.~m} in Figure~\ref{fig:pathLegs} that makes a lift request after the initial path has been computed). 
In the presence of other passengers on the path, the AS 
will decide whether or not to offer a lift to them, based on the objectives of the passengers already on board. Accepting new passengers lead AS to dynamically update their path (i.e., by adding \texttt{leg~5} and \texttt{leg~6} in place of \texttt{leg~1} in Figure~\ref{fig:pathLegs} when \texttt{Pass.~m} is accepted on board). Moreover, an optimized decision process supporting the dynamic addition of passengers 
would be beneficial for both AS, which can reach their full occupancy, and for the passengers, who would share the travel costs. Of course, not all passengers on the path can be accepted if it means that having them on board causes a degradation of the travel duration. The latter can be particularly relevant for those passengers on their way to work. 

The proposed solution is supported by a set of algorithms: $(i)$~to compute an initial path on the base of the first accepted passenger(s), $(ii)$~to compute the various legs of the AS path, with the related costs, 
$(iii)$~to dynamically add passengers on the way, and $(iv)$~to update the costs each time a new passenger(s) is added to a travelling AS~\footnote{For a matter of space, we only present the algorithm to manage self-organized AS, when new passengers should be added (see \textbf{Algorithm 1}). The complete set of algorithms can be found at the following link: \url{https://bit.ly/2M9w9n8}.}.
Algorithm~1 presents the algorithm to manage self-organized AS, when new passengers should be added, i.e., the core of the approach.

Each time a shuttle finds passengers on the road, it will set an initial list of new \textit{potential passengers} with their destinations. To decide about their acceptance on board, a previous analysis needs to be performed. Consequently, this list will be first filtered by the \textit{angle} between the passengers (both on board and new) destinations and the current location of the car. Each new passenger whose computed angle is greater than 20° would be removed from the list, as they are considered too far from the AS current location and serving them would imply a significant negative effect on the time of travel. For instance, Figure~\ref{fig:pathLegs} shows that \texttt{Pass.~m} has been accepted since the angle between his/her current position and the position of the AS is equal to 20°.
Indeed, the greater the angle is, the greater the travel time will be. Thus, the angle degree must be kept under a certain upper bound value of 20° that we get after running several experiments showing that for values higher than 20° a constant worsening of performance could be observed.
After, the list will be filtered by the stops already visited by the AS. 
In fact, in order to avoid AS to significantly increase its total path, 
those passengers whose destinations correspond to stops that have already been visited are not accepted for the lift. This helps to avoid AS to make a non-optimized route by revisiting already visited stops on its path to the final destination.

After the initial checks, if there are still passengers in the list, a loop over the new potential destinations, namely $d$, starts in order to further understand whether it is possible to include them as future stops or if the passengers should either wait for another AS or go by their own to find an alternative solution to arrive to their destinations. This loop is the core part of Algorithm~1 (lines~\textbf{5--40}). 
First, the last index, namely $max\_index$ of the \textit{targets} list is acquired 
(line \textbf{7}), then a check is made over the length of the passengers that will be dropped off by the AS 
at destination $d$: only the number of passengers needed to fill the remaining available seats are considered. If there are no passengers left or, if there are no available seats left (lines \textbf{8-9}), the cycle is finished. 
Otherwise, it is checked whether or not the destination of  passengers is already among the next stops of the AS. If that is the case, the passengers are added on board (lines \textbf{11-12}). Otherwise, a loop starts (lines \textbf{14-38}) over the next stops in order to insert the destination at the appropriate index. 
The above implies several calculations. The path between the current target and the next target, namely $t2n$, and the path between the current target and the destination $d$, namely $t2d$, are computed (lines~\textbf{15-20}). If they are defined for the first time (line~\textbf{15}), the current target will correspond to the $new\_origin$ (lines~\textbf{16-17}), otherwise it will be the destination at index $i-1$ (lines~\textbf{19-20}). This distinction is necessary since the origin is not in the list of targets, but only in the list of global stops. 
The length of these two paths are compared (line~\textbf{21}) and the best option is chosen accordingly. 
Specifically, if the length to the destination $t2d$ is shorter than the one to the next stop $t2n$, the path from the destination $d$ to the next stop is computed (line \textbf{22}). Before the addition of the passengers and their destinations, some additional checks about the very first passenger are needed, as for example, if he/she is going to work (line \textbf{24}). 
An approximation about the time needed to reach his/her working place considering the path variation due to the addition of new passengers is then computed (line \textbf{25}). 
Thereafter, it is checked whether the time variation is affordable, in order avoid lengthening the path unnecessarily. This information is stored in the $check$ variable (line \textbf{26}) that is used to decide whether adding or not the new passengers (lines \textbf{27-28}). At this point, if necessary, the AS current path is recomputed, based on the updated list of stops.
If the length to the destination $t2d$ is longer than the one to the next stop $t2n$, the counter $i$ will be incremented and therefore, the loop will start again.  Eventually, when the value of $i$ corresponds to the last index of the $targets$ list, the destination will be added to the $targets$ and to the $stops$ as the last stop (lines \textbf{33-37}). 

After each cycle of the loop over the destinations, the passengers will be removed from the list of potential new passengers (line \textbf{40}). 
Lastly, if there have not been any additional passengers at the current location, i.e., $tot\_added\_passengers = 0$, the $new\_origin$, will be removed from the global list of stops, unless it is the origin of the travel (lines \textbf{41-43}).

After the actual new passengers have been identified, 
the following further steps are performed. The current position of each passenger (i.e. the origin of their travel) is added to the list of the AS' stops before the next stop. 
Moreover, the path from the AS current location to the destination of the first passenger who boarded the vehicle (and therefore, whose destination coincides with the final target of the AS), is computed again in order to include the new stops. 
The time to the destination is also updated for each passenger to therefore, make sure they are not running too late and, eventually, the number of remaining seats is computed.

In terms of temporal complexity, the decentralized algorithm just described is \textit{subquadratic} with $O(n*m)$. In particular, $n$ is the number of passengers that must be managed by an AS while $m$ is the number of stops that the same AS must handle. Moreover, since each AS can serve at most 12 passengers at a time, and new passengers cannot be taken on board if there are no more seats, in the worst case $m$ will be equal to 11.

\begin{algorithm}[!ht]
\footnotesize{
\KwData{ $P$; $targets$;$original$;}{}{}
\SetKwData{p}{$p_1$}\SetKwData{pp}{$p_2$} \SetKwData{on}{On\_board}
\SetKwFunction{anglerem}{rem\_angle}
\SetKwFunction{upcosts}{Update\_Cost\_Legs}
\SetKwFunction{changep}{Change\_Path}
\SetKwFunction{addp}{Add\_Passengers}
\SetKwInOut{Input}{input}\SetKwInOut{Output}{output}
\KwResult{$p \in Passengers$;  updated $stops$; updated $targets$; updated $cost\_legs$;}
    tot\_added\_passengers $\leftarrow$ 0\;
    up\_costs\_pass $\leftarrow$ \textbf{false}\;
    added $\leftarrow$ \textbf{false}\;
    open\_seats $\leftarrow$ max\_pass - $|$passengers$|$\;
    
    \ForEach{d $\in$ destinations}{
        i $\leftarrow$ 0\;
        max\_index $\leftarrow$ $|$targets$|$-1\;
        \eIf{open\_seats $=$ 0 \textbf{or} $|$destinations[d]$| =$ 0}{
            \textbf{break}\;
        }{
            \eIf{d $\in$ targets}{
                \addp\;
            }{
                \While{added $=$ \textbf{false}}{
                    \If{i $=$ 1}{
                        t2d $\leftarrow$ path(new\_origin, d)\;
                        t2n $\leftarrow$ path(new\_origin, targets[i])\;
                    }\uElseIf{i $<$ max\_index}{
                        t2d $\leftarrow$ path(targets[i-1], d)\; 
                        t2n $\leftarrow$ path(targets[i-1], targets[i])\;
                        \tcc{where $i-1$ is the index of the current stop}
                    }
                    \If{$|$t2d$| < |$t2n$|$}{
                        d2t $\leftarrow$ path(d, targets[i])\;
                        check $\leftarrow$ \textbf{true}\;
                        \If{'work' \textbf{in} first.next\_state}{
                            change $\leftarrow$ t\_a(original) - t\_a(t2n) + time\_a(t2d) + t\_a(d2n)\;
                            check $\leftarrow$ (change$<$(t\_a(original)$\times$3/2)) \textbf{and} (av\_time$\geq$change)\;
                        }
                        \eIf{check}{
                            \addp\;
                            \If{i $=$ 1}{
                                \changep\;
                            }
                        }{
                            \textbf{break}\;
                        }
                    }
                    \If{i $=$ max\_index  \textbf{and not} added}{
                        targets $\leftarrow$ targets + d\;
                        stops $\leftarrow$ stops + d\;
                        \addp\;
                        as\_last $\leftarrow$ \textbf{true}\;
                    }
                    \textbf{break}\;
                }
                i $\leftarrow$ i+1\;
            }
            $P \leftarrow P$ - destinations[d]\;
        }
    }
    \If{tot\_added\_passengers $=$ 0}{
        \If{stops index\_of new\_origin $\neq$ 0}{
            stops $\leftarrow$ stops - new\_origin\;
        }
    }
 \caption{\footnotesize{Decentralised algorithm for adding new passengers after the first passenger or group.}}
 }
\end{algorithm}
As introduced above, the real-time dynamic ride-sharing is built on the idea of sharing a vehicle among individual travellers for a trip, while splitting the travel costs. 
This allows users to reduce the costs, although it may potentially occur at the expense of passenger's convenience.
Computing the final cost for each passenger is not easy when passengers are dynamically boarding and leaving cars. Specifically, the total path of each AS corresponds to a sequence of legs, which is continuously modified due to the addition of passengers on board. Each passenger must contribute only to the costs of those legs for which he/she was on board, independently on the entire path that the AS covers. 
Accordingly, in our approach the travel cost is calculated based on the computation of the costs of legs shown by $cost_{leg}$ in equation~(\ref{eq:costLeg}), where the length of the leg and the cost per km are considered.
\begin{equation}
    \label{eq:costLeg}
    cost_{leg} = \frac{length(leg)}{1000} \times cost_{Km}
\end{equation}

Then, for each AS, the travel cost corresponding to the entire path it covered is calculated by the sum of the costs of the legs in the path, $cost_{path}$ in equation~(\ref{eq:costPath}).
\begin{equation}
    \label{eq:costPath}
    cost_{path} = \sum_{i=1}^n cost_{leg_{i}}
\end{equation}

Eventually, to compute the cost that each passenger must pay on a given leg for which he/she was on board, namely $cost_{leg_{pass}}$, the $cost_{leg}$ is divided by the number of passengers for that leg, as implied by equation~(\ref{eq:costLegPass}). 
\begin{equation}
    \label{eq:costLegPass}
    cost_{leg_{pass}} = \frac{cost_{leg}}{|passengers\_on\_board|}
\end{equation}
Of course, the total cost for each passenger will be given by the sum of the $cost_{leg_{pass}}$ of each leg he/she has to contribute.

%% file: conclusion.tex
Planning for traffic patterns that will incorporate intelligent vehicles and deal with the inherent uncertainty is a daunting task. It requires the ability to analyze the effects of systems that do not exist yet, while at the same time committing to the building of transportation infrastructures envisioned to last a long term. Hence, the ability for city planners to plan for these potential futures is critical.
The scientific literature has already investigated the potential impacts of dynamic ride-sharing with automated vehicles. 
Instead, this paper has described a system that enables city mobility planners to explore the integration of AS, and the potential of tackling the uncertainty underlying, using a decentralized control. 
Expectedly, the integration of AS has significant potential to improve traffic capacity and efficiency of mobility systems, assuming different trade-offs.

As seen in more detail in the experiments, a mix of traditional and intelligent vehicles can also be analyzed using the proposed framework. 
A final goal is to additionally develop the framework into a further robust, off-the-shelf tool that planners can use to optimize novel technologies in towns. 

The insertion of cooperative behaviours among AS have great potential in the framework, to therefore help when AS face unexpected situations such as breakdowns, heavy traffic and congestion. 
We are experimenting with new cooperative behaviours to face such problems. 
As such, future directions include the implementation of adaptation strategies on top of those self-organization provided to handle adaptation needs, such as, i) the dynamic rescheduling of the available AS in cases in which there are passengers who cannot longer be reached by the assigned car; ii) the dynamic re-grouping of passengers of a faulty car, among the closest AS around. 

The behaviour of systems with AS is emerging and uncertain, and can therefore surprise the stakeholders~\cite{Bencomo_surprise2014}. Therefore, we also plan to add explanation capabilities about the decision-making and impact of self-organized and adaptive cooperation of AS to the framework~\cite{DBLP:conf/models/Garcia-Dominguez19, sawyer_requirements-aware_2010, DBLP:journals/tcci/WelshBSW14,Bucchiarone19}. 
One of the potential benefits of creating this framework is that it allows planners to \enquote{score} a configuration along multiple parameters, including cost, traveller waiting time, carbon production, etc. Not only it provides flexibility to the planner, but also provides a sound basis for integrating machine learning into the planning process. Using a Reinforcement Learning (RL) approach~\cite{sutton2018reinforcement}, the outputs of the simulation can be used as a cost function that can train a system to find local (and potentially global) optima in a complex, multidimensional environment. A next step in this research effort is to integrate RL approaches into the traffic planning process to work according to the different \enquote{best} options according to given criteria. 
A goal is to  have a planner outlining a region on a map to select options about the population, and let the RL decision-making calculate the different \enquote{best} options according to the criteria. 
Providing a RL based evaluation allows for the assessment of a broad range of possibilities to be further explored by human planners. For example, human planners would take care of edge cases such as those where passengers reject a vehicle's proposal, handling non-standard cargoes, or providing transportation services that need special coordination with other stakeholders such as parents. 

Eventually, since the connection between autonomous vehicle technologies and estimated changes in energy consumption is receiving a great deal of attention \cite{ROSS2017}, our plan it to extend our framework to deal with mobility scenarios including autonomous electric shuttles and simulate different situations with the objective to optimize energy resources in a city.